\documentclass[aps,prl,twocolumn,superscriptaddress,floatfix,longbibliography,nofootinbib]{revtex4-2}

\usepackage{booktabs}
\usepackage{multirow}
\usepackage{graphicx,color}
\usepackage{duckuments}
\usepackage{bm}
\usepackage{color}
\usepackage[english]{babel}
\usepackage{float}
\usepackage{mathrsfs}
\usepackage{verbatim}
\usepackage{amsmath}
\usepackage{hyperref}
\usepackage{physics}
\usepackage{tikz}
\usepackage{etoolbox}
\usepackage[normalem]{ulem}
\usepackage{soul}
\usepackage{todonotes}
\usepackage{microtype}
\usepackage{bibunits}
\defaultbibliography{references.bib}
\defaultbibliographystyle{apsrev4-2}

\usepackage[T1]{fontenc}
\usepackage[utf8]{inputenc}

\usepackage{amsthm}
\usepackage{amssymb}
\usepackage{enumitem}

\def\figureautorefname~#1\null{{\textcolor{black}{Fig.}~}#1\null}
\def\sectionautorefname~#1\null{{\textcolor{black}{Sec.}~}#1\null}
\def\tableautorefname~#1\null{{\textcolor{black}{Tab.}~}#1\null}

\def\equationautorefname~#1\null{{\textcolor{black}{Eq.}~}(#1)\null}

\definecolor{max}{HTML}{03A678}
\definecolor{maciek}{HTML}{161226}
\definecolor{roy}{HTML}{0312A8}

\renewcommand{\selectlanguage}[1]{}

\date{\today}
\begin{document}
\begin{bibunit}

\title{Autoencoder-based analytic continuation method \\
for strongly correlated quantum systems}
\author{Maksymilian Kliczkowski}
\affiliation{Institute of Theoretical Physics, Wrocław University of Science and Technology, 50-370 Wrocław, Poland\looseness=-3}
\author{Lauren Keyes}
\affiliation{Department of Physics, The Ohio State University, Columbus, Ohio 43210, USA\looseness=-3}
\author{Sayantan Roy}
\affiliation{Department of Physics, The Ohio State University, Columbus, Ohio 43210, USA\looseness=-3}
\author{Thereza~Paiva}
\affiliation{Instituto de F\'isica, Universidade Federal do Rio de Janeiro Cx.P. 68.528, 21941-972 Rio de Janeiro RJ, Brazil}
\author{Mohit Randeria}
\affiliation{Department of Physics, The Ohio State University, Columbus, Ohio 43210, USA\looseness=-3}
\author{Nandini Trivedi}
\affiliation{Department of Physics, The Ohio State University, Columbus, Ohio 43210, USA\looseness=-3}
\author{Maciej M. Ma\'ska}
\affiliation{Institute of Theoretical Physics, Wrocław University of Science and Technology, 50-370 Wrocław, Poland\looseness=-3}
\begin{abstract}
The single particle Green's function provides valuable information on the momentum and energy-resolved spectral properties for a strongly correlated system. In large-scale numerical calculations using quantum Monte Carlo (QMC), dynamical mean field theory (DMFT), including cluster-DMFT, one usually obtains the Green's function in imaginary-time $G(\tau)$. The process of inverting a Laplace transform to obtain the spectral function $A(\omega)$ in real-frequency is an ill-posed problem and forms the core of the analytic continuation problem. 
In this Letter, we propose to use a completely unsupervised autoencoder-type neural network to solve the analytic continuation problem. We introduce an encoder-decoder approach that, together with only minor physical assumptions, can extract a high-quality frequency response from the imaginary time domain. With a deeply tunable architecture, this method can, in principle, locate sharp features of spectral functions that might normally be lost using already well-established methods, such as maximum entropy (MaxEnt) methods. We demonstrate the strength of the autoencoder approach by applying it to QMC results of $G(\tau)$ for a single-band Hubbard model. The proposed method is general and can also be applied to other ill-posed inverse problems. 
\end{abstract}

\maketitle

\noindent \ul{\it Introduction:} Solving ill-posed inverse problems is critical across scientific disciplines, where the reconstruction of functions from indirect or noisy observations presents inherent challenges \cite{kabanikhin_inverse_2011, tanaka_inverse_2021, tarantola_inverse_1987}. Prominent ill-posed inverse problems include, e.g., recovering the signal from convoluted or blurred versions~\cite{jacquelin_force_2003, asim_blind_2019, adke_detection_2021, hormann_consistently_2022}, reconstructing a function from its Laplace transform~\cite{de_hoog_improved_1982, van_iseghem_laplace_1987, kwok_algorithm_1989, gzyl_maxentropic_1995}, determining object properties from scattered waves~\cite{scotti_shape_1995, scotti_shape_1996}, finding the initial temperature distribution within a material based on temperature measurements at its surface~\cite{kant_determination_2016}, or reconstructing a function from a finite set of data. In this Letter, we propose an autoencoder-type neural network as a universal tool for solving ill-posed inverse problems. To demonstrate its effectiveness, we apply it to perform analytic continuation, transforming imaginary-time Quantum Monte Carlo (QMC) data to real frequency. Given its universality, the same scheme can address various ill-posed inverse problems, e.g., it can be used instead of the Lucy-Richardson iterative deconvolution technique \cite{lucy_iterative_1974, richardson_bayesian-based_1972} for spectroscopy data.

Frequency-dependent functions, transformed from imaginary-time correlators, such as Green's functions [referred to as $G(\tau)$], describe the real-time propagation of particles and excitations in a system. A spectral function (SF), $A(\omega)$, can be readily transformed into the corresponding $G(\tau)$ through a straightforward analytic convolution, otherwise known as the Fredholm integral of the first kind~\cite{polianin_handbook_1998, zappala_neural_2023, guan_solving_2022}, 
\begin{equation}
G(\tau)=\int_{-\infty}^{\infty} {\cal K}(\tau,\omega)A(\omega)d\omega,
\label{eq:direct}
\end{equation}
where for fermions the kernel ${\cal K}$ is given by
\begin{equation}
{\cal K}(\tau,\omega)=-\dfrac{e^{-\omega\tau}}{1+e^{-\beta\omega}}
\label{eq:kernel}
\end{equation}
for $0 \leq \tau \leq \beta = 1/T$, the inverse temperature.
Although it is easy to calculate the integral in Eq.~\eqref{eq:direct}, the reverse transformation is ill-posed, exemplifying an inverse problem. It is also highly susceptible to statistical and numerical errors. Even slight fluctuations in the input can result in significant discrepancies in the final result~\cite{zhang_training_2022}. This makes the problem of analytic continuation extremely challenging, and due to its importance, for, e.g., quantum many-body physics, tremendous effort is being invested into its solution~\cite{shao_progress_2023}.

By discretizing imaginary time $\tau_i$ and frequencies $\omega_j$ and introducing notation $G_i\equiv G(\tau_i), A_j \equiv A(\omega_j)$, Eq.~\eqref{eq:direct} can be rewritten in a matrix form,  
\begin{equation}
    \vec{G}={\cal K}\vec{A},
    \label{eq:direct1}
\end{equation}
where $\vec{G}=(G_1,\ldots,G_N)$,  $\vec{A}=(A_1,\ldots,A_M)$, and ${\cal K}$ is a $N\times M$ kernel matrix. The direct approach to determining $A(\omega)$ would be to calculate a generalized inverse of Eq.~(\ref{eq:direct1})~\cite{rao_generalized_1971, ben-israel_generalized_2003}. However, the  difficulties mentioned above usually render this simple idea unusable.
Therefore, different methods have been proposed, including Pad\'e approximants~\cite{schott_analytic_2016}, stochastic analytic continuation~\cite{sandvik_stochastic_1998, shao_progress_2023}, sparse modeling~\cite{otsuki_sparse_2017, yoshimi_spm_2019, motoyama_robust_2022}, spectrum averaging~\cite{lotsch_average_1991}, stochastic pole expansion~\cite{huang_stochastic_2023}, genetic algorithms~\cite{vitali_ab_2010} and maximum entropy (MaxEnt) methods~\cite{pavarini_correlated_2012, levy_implementation_2017}, with the last being the most widely used. Significant effort has been put into their enhancement, but the search for a fully reliable method continues. Recently, a new approach has been introduced using the ``Nevanlinna'' structure of Green's function~\cite{fei_nevanlinna_2021}. This approach undertakes interpolation rather than searching for a fit that matches multiple predefined conditions. A robust extension of this method, called PES, has been proposed in Ref.~\cite{huang_robust_2023}. 

In this Letter, we demonstrate the efficiency and accuracy of a different strategy, involving rapidly advancing Machine Learning (ML) techniques to tackle this issue. We compare it with MaxEnt [details can be found in the Supplemental Material (SM)]. Neural Networks (NN) have already been proposed to obtain $A(\omega)$ from $G(\tau)$ \cite{fournier_artificial_2020,yoon_analytic_2018,zhang_training_2022,yao_noise_2022}, producing remarkable results compared to MaxEnt. These approaches were based on a popular type of ML, {\it supervised learning}, where a NN is trained to provide an expected output for a given input. Since supervised learning requires labeled data, a large number of SFs must first be artificially generated in a physically meaningful way. These SFs are then the ``labels'', i.e., the expected outputs. The corresponding $G(\tau)$'s are calculated according to Eq.~\eqref{eq:direct} and pairs $\left[G(\tau), A(\omega)\right]$ are used to train the NN. Because this type of NN is trained on ``artificial'' SFs, it does not necessarily perform well for real QMC data. Our approach incorporates supervised learning as a first step, but ultimately trains on real QMC data.  

\vskip 2mm

\noindent \ul{\it Autoencoder approach:} We propose an approach which trains using $G(\tau)$'s obtained directly from QMC. In this case, we lack the knowledge of the exact corresponding $A(\omega)$, making it inaccessible to supervised learning. To overcome this, we propose an {\it unsupervised} method based on an {\it autoencoder} (AE) NN. Starting with initial weights (described in the \underline{\it Pretraining} section), the NN calculates $A(\omega)$ from the original $G(\tau)$, and then uses the straightforward analytic formula \eqref{eq:direct} to calculate $G'(\tau)$. The NN is then trained to minimize the difference between $G(\tau)$ and $G'(\tau)$, without requiring any prior knowledge of $A(\omega)$. Equation \eqref{eq:autoencoder} illustrates this idea,
\begin{equation}
\vec{G} \hspace*{2mm}\raisebox{-0.7mm}{\includegraphics*[width=6mm]{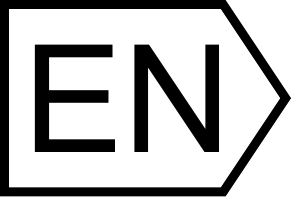}}\hspace*{1.5mm} \vec{A} \hspace*{2mm}\raisebox{-0.7mm}{\includegraphics*[width=6mm]{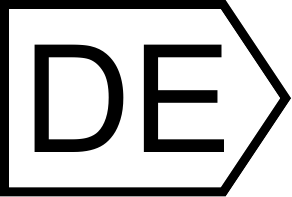}}\hspace*{2mm} \vec{G}',
\label{eq:autoencoder}
\end{equation}
in which \raisebox{-0.7mm}{\includegraphics*[width=5mm]{FinalFig/EN.pdf}} and \raisebox{-0.7mm}{\includegraphics*[width=5mm]{FinalFig/DE.pdf}} represent the {\it encoder} and {\it decoder} parts of the NN, respectively. $\vec{A}$ becomes the latent vector of the AE. The encoder structure is described in the SM. This part of the NN performs the inverse transformation $\vec{A}=f(\vec{G})$. This is possible due to the {\it Universal Approximation Theorem}, asserting that NN of sufficient size can accurately approximate any piecewise continuous function~\cite{hornik_multilayer_1989,siegelmann_computational_1992,barron_universal_1993}. The decoder is a single layer that performs a  matrix multiplication, as given by Eq.~\eqref{eq:direct1}. The only trainable layers are in the encoder. The weights and biases are adjusted to make the reconstructed Green's function $\vec{G}'$ as close to the original Green's function $\vec{G}$ as possible. Since the decoder part performs a well-posed forward transformation \eqref{eq:direct1}, the encoder is trained to perform the inverse transformation with the SF $\vec{A}$ as its output. The natural choice for the loss function, i.e., the function minimized during training, is a squared distance between $\vec{G}$ and $\vec{G}'$~\cite{actual_footnote},
\begin{equation}
    \chi^2=||\vec{G}-\vec{G}'||^2.
\label{eq:chi2} 
\end{equation}
Unfortunately, similarly to the MaxEnt, minimization of $\chi^2$ suffers from the lack of uniqueness. One possibility is to introduce, as is done in MaxEnt, an additional entropy term, which favors the similarity of $A(\omega)$ to a chosen default model. However, this is flawed with the ambiguity in choosing the default model and the competing importance of $\chi^2$ with the entropy term. NNs are well suited for the application of methods that help to tackle this ambiguity, such as imposing sum rules or smoothness of the SF. While these techniques are easier to implement in NNs than in MaxEnt, they only partially remove the solution ambiguity. Therefore, in this Letter, we propose a different approach that we describe below. 

\vskip 2mm

\noindent\ul{\it Pretraining:} 
Since the forward problem \eqref{eq:direct} at fixed $\beta$ does not depend on a particular physical model, the inverse transformation should be unique and independent of the nature of Green's functions. However, this would require infinite numerical precision. Moreover, since our approach represents the transformation as a NN, and too many neurons would make training infeasible, we have to work with an approximate form of the inverse transformation. To make it highly efficient and accurate, we propose a two-stage procedure: 
\begin{enumerate}
\item Firstly, the \,\raisebox{-0.7mm}{\includegraphics*[width=5mm]{FinalFig/EN.pdf}} learns general characteristics typical for SFs, and 
\item Secondly, it is tuned to represent the transformation for particular $G(\tau)$'s obtained in QMC. 
\end{enumerate}
\begin{figure}[!t]
\centerline{\includegraphics[width=0.45\textwidth]{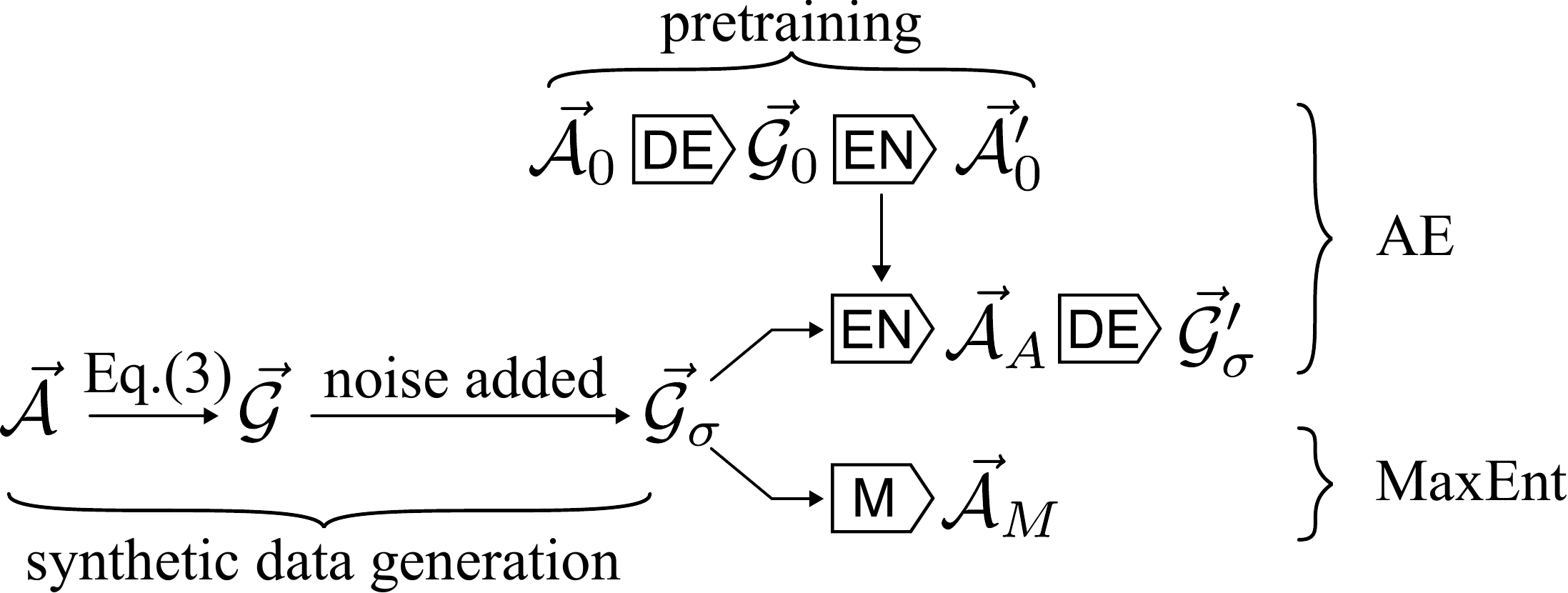}}
\caption{Illustration of the way we test the robustness of our approach against statistical errors. \raisebox{-0.7mm}{\includegraphics*[width=5mm]{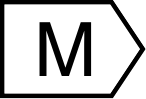}} represents the MaxEnt procedure; $\vec{\cal A}_A$ [$\vec{\cal A}_M$] denote the SFs obtained within the AE [MaxEnt] methods. The part denoted as ``autoencoder'' includes the pretraining and the actual AE training. We use index ``0'' for the SFs and Green's functions
used at pretraining stage to distinguish them from SFs used in the actual error estimation procedure. Details are given in the text. 
\label{fig:testing_scheme}}
\end{figure}
In the first stage, we follow the method of Refs. \cite{fournier_artificial_2020,yoon_analytic_2018,zhang_training_2022,yao_noise_2022} and use ``artificial'' SFs $\vec{\cal A}_0$ modeled as a sum of random number of Gaussian peaks. For each SF, we calculate the corresponding Green's function $\vec{\cal G}_0$ according to Eq.~\eqref{eq:direct1}. Then, we use a large set of pairs $(\vec{\cal G}_0^n, \vec{\cal A}_0^n)$ to train \raisebox{-0.7mm}{\includegraphics*[width=5mm]{FinalFig/EN.pdf}}. We use calligraphic letters to distinguish the ``artificial'' SFs and the corresponding Green's functions from the ones obtained in QMC simulations. Formally, by analogy to Eq.~\eqref{eq:autoencoder}, the procedure described above can be illustrated as
\begin{equation}
    \vec{\cal A}_0\hspace*{2mm}\raisebox{-0.7mm}{\includegraphics*[width=6mm]{FinalFig/DE.pdf}}\hspace*{1.5mm} \vec{\cal G}_0\hspace*{2mm}\raisebox{-0.7mm}{\includegraphics*[width=6mm]{FinalFig/EN.pdf}}\hspace*{1.5mm} \vec{\cal A}_o',
\end{equation}
where \raisebox{-0.7mm}{\includegraphics*[width=5mm]{FinalFig/EN.pdf}} now plays the role of decoder and is trained to minimize~\cite{since_footnote}
\begin{equation}
\eta^2=||\vec{\cal A}_0-\vec{\cal A}_0'||^2.
\end{equation}
However, since \raisebox{-0.7mm}{\includegraphics*[width=5mm]{FinalFig/DE.pdf}} is given by Eq.~\eqref{eq:direct1} and we know $\vec{\cal A}$, this is a standard problem of supervised learning, which usually has a well-defined unique solution. 
We call this {\it pretraining}, since at this stage the NN does not learn how to transform real QMC data. Instead, the network learns general features of SFs. This step is also useful for addressing the ``curse of dimensionality'' \cite{bellman_dynamic_1957}. Namely, the {\it pretrained} weights and biases of \raisebox{-0.7mm}{\includegraphics*[width=5mm]{FinalFig/EN.pdf}} provide suitable initial conditions for the second stage to prevent backpropagation from becoming trapped in a suboptimal local minimum. Thus, in actual training with QMC $G(\tau)$'s according to Eq.~\eqref{eq:autoencoder}, we only adjust already pretrained weights and biases of \raisebox{-0.7mm}{\includegraphics*[width=5mm]{FinalFig/EN.pdf}}. 
\begin{figure}[!ht]
\centerline{\includegraphics[width=0.47\textwidth]{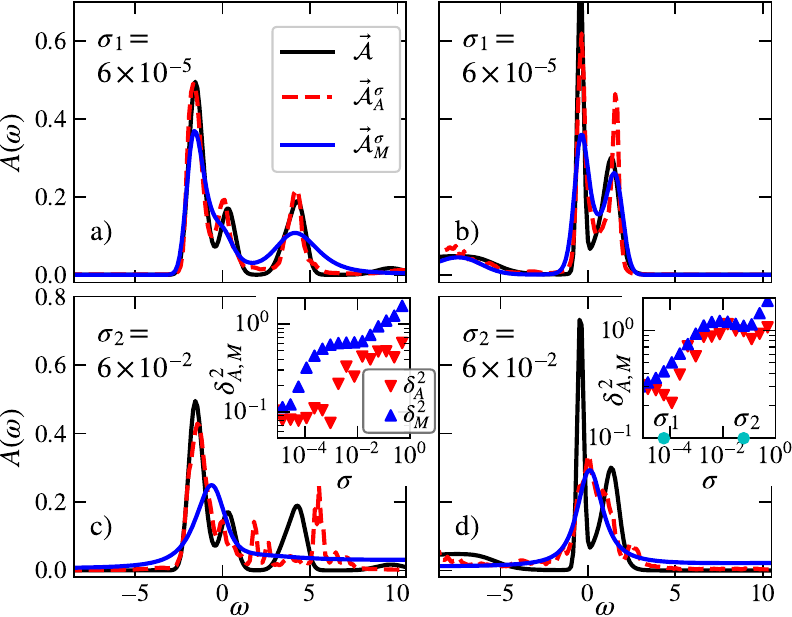}}
\caption{(a)-(d) Spectral functions (SFs) $\vec{\cal A}_{A, M}$ predicted by auto-encoder(AE) (red dash-dotted line) and MaxEnt (blue dashed line) from Green's functions ${\cal G}$ calculated for artificial SFs. True $\vec{\mathcal{A}}_0$ is marked with a solid black line. Results for two different SFs [one in (a) and (c), another in (b) and (d)] and for two values of the noise $\sigma$ imposed on artificial Green's functions [$\sigma_1=6\times 10^{-5}$ in (a) and (b), $\sigma_2=6\times 10^{-2}$ in (c) and (d)].
The insets show $\sigma$-dependence of the reconstruction errors $\delta_{A, M}^2$ [cf. Eq.~\eqref{eq:deltas}] corresponding to the SF. The cyan circles indicate the values of $\sigma_1$ and $\sigma_2$. \label{fig:errors}}
\end{figure}

In the MaxEnt approach, prior knowledge of the spectra is introduced by defining the default model, which is used to select one of the many solutions to minimization of $\chi^2$ in Eq.~\eqref{eq:chi2}. In the present approach, the pretraining plays a similar role. Here instead, it forces the SFs to be close to a realistic multi-peak structure, as opposed to simple default models typically used in MaxEnt. Once acquired, the weights can be employed subsequently without the need for recalculation.

While pretraining is the main way to ensure unambiguity of the solution, it is easy to apply other techniques to improve the quality of the SFs obtained. We impose a limited number of physical assumptions about the SFs, such as their normalization and non-negativity. The flexibility of NNs enables the incorporation of any desired regularization. For instance, we can introduce the entropy term, typically maximized in the MaxEnt procedure, as a separate layer within the NN framework. We explore other possibilities, including simple $L1$ and $L2$ penalties (absolute and squared values of weights, respectively), as well as more sophisticated regularizations based on first- and second-derivative constraints to ensure the SF's smoothness. We further follow the idea of White \cite{white_spectral_1991} to incorporate constraints put by spectral moments that are known analytically. Using the flexibility and power of NNs, we demonstrate the ability to tackle the challenges posed by statistical noise and the absence of prior knowledge of the SF in a more universal manner. 
\vskip 2mm

\noindent \ul{\it Comparing performance: Synthetic data.} To demonstrate the advantages of the proposed method, we need measures to evaluate its efficiency and accuracy. There is no obvious way to assess the AE's accuracy. The true SFs corresponding to $G(\tau)$'s produced by QMC are {\it a priori} unknown, and thus we cannot determine whether the result produced by the AE or by MaxEnt is more accurate. To overcome this difficulty, we compare the performance of the AE and MaxEnt on synthetic noisy Green's functions, for which the exact SF is known. Namely, we generate ``artificial'' SFs of the same nature as those used in pretraining. Subsequently, we use Eq.~\eqref{eq:direct} to calculate corresponding $\mathcal{G}(\tau)$'s. Finally, we add noise $\sigma$ that mimics averaging over different lengths of QMC runs. Various magnitudes of noise $\sigma$ are added to $\mathcal{G}(\tau)$ to produce a set $\vec{\cal G}_\sigma$ (for details, see SM). The resulting set of $\vec{\cal G}_\sigma$'s is used to perform the second training stage of the NN [Eq.~\eqref{eq:autoencoder}]. Once training is finished, we test the AE [MaxEnt] by applying it on sets $\vec{\cal G}_\sigma$ independent of those used to train the NN and produce SFs $\vec{\cal A}_A$ [$\vec{\cal A}_M$]. This process is illustrated in Fig.~\ref{fig:testing_scheme}. Since we know the original SF $\vec{\cal A}$, we explicitly calculate the errors introduced by both the AE and MaxEnt. We define
\begin{equation}
\delta_A^2=||\vec{\cal A}-\vec{\cal A}_A||^2,\quad \delta_M^2=||\vec{\cal A}-\vec{\cal A}_M||^2,
\label{eq:deltas}
\end{equation}
where $\delta_A^2$ and $\delta_M^2$ measure the AE and MaxEnt errors, respectively.
\begin{figure}[!ht]
\centerline{\includegraphics[width=0.46\textwidth]{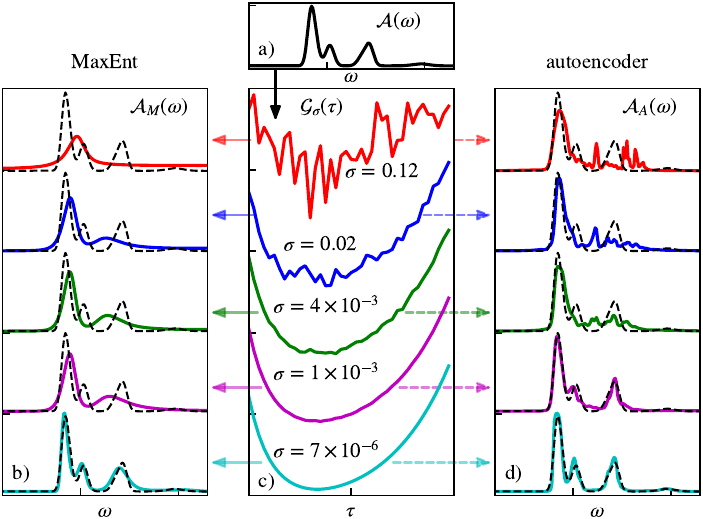}}
\caption{
(a) An example of ``artificial'' SF ${\cal A}(\omega)$, used to calculate, according to Eq.~\eqref{eq:direct1}, the corresponding Green's function ${\cal G}(\tau)$. This is represented by a vertical black arrow. Noisy ${\cal G}_\sigma(\tau)$'s are shown in panel (c). In panel (d) [(b)] the \raisebox{-0.7mm}{\includegraphics*[width=5mm]{FinalFig/EN.pdf}} [MaxEnt] is used to calculate SFs ${\cal A}_A(\omega)$ [${\cal A}_M(\omega)$]. This is represented by dashed right [solid left] arrows. The original SF is shown with a black dashed line in both panels.
\label{fig:greensandspectrals1}}
\end{figure}

Panels (a)-(d) of Fig.~\ref{fig:errors} show examples of SFs obtained with the AE and MaxEnt for two different noise magnitudes, $\sigma_1$ and $\sigma_2$. Figs.~\ref{fig:errors} (e) and (f) show the corresponding dependence of $\delta_A^2$ and $\delta_M^2$ on $\sigma$. It can be seen that $\delta_A^2$ is smaller than or equal to $\delta_M^2$ for all values of $\sigma$, indicating that shorter QMC runs can be used with the AE method to produce SFs with equal or greater accuracy than those produced by MaxEnt. We further demonstrate the impact of noise in Fig.~\ref{fig:greensandspectrals1}. In panel (d) [(b)] we show the SFs $\mathcal{A}_A(\omega)[\mathcal{A}_M(\omega)]$ obtained by using AE [MaxEnt] on the $G_\sigma(\tau)$'s given in panel (c) for various $\sigma$. Comparison of $\mathcal{A}_{A,M}(\omega)$ with the original SF $\mathcal{A}(\omega)$ [Fig.~\ref{fig:greensandspectrals1}(a)] directly highlights the robustness of the AE to noise. By evaluating the AE's performance across $\sigma$, we examine the resulting SF's susceptibility to statistical errors. We show that, having trained on a broad set of noisy Green's functions, the AE adeptly captures crucial SF characteristics compared to simple averaging. This ability is the main reason why AEs are frequently used in image denoising \cite{vincent_stacked_2010}. During training, in addition to learning to transform data to the frequency domain, the AE learns to distinguish between inherent statistical errors in QMC and key features that define the quantum system. 
\vskip 2mm

\noindent \ul{\it Comparing performance: QMC data.} Next, we test our approach on real QMC data. To do this, we use single-particle imaginary-time Green's functions, obtainable in Determinant Quantum Monte Carlo (DQMC) simulation, which are defined as
\begin{equation}\label{eq:greens}
    G_{\vec{k}}(\tau) = -\sum_\sigma\langle \hat{c}_{\vec{k}\sigma}(\tau)\hat{c}_{\vec{k}\sigma}^\dag (0) \rangle,
\end{equation}
where $\vec{k}$ stands for the particle momentum vector and $\hat{c}_{\vec{k}\sigma}^\dag(\tau), \hat{c}_{\vec{k}\sigma}(\tau)$ denote fermionic creation and annihilation operators at imaginary time $\tau$. We study the Fermi-Hubbard model (details in SM). The DQMC method is limited to intermediate temperatures so as to keep the average sign of the determinants away from zero ~\cite{troyer_computational_2005, hangleiter_easing_2020, pan_sign_2022}. In this Letter we focus on parameter regimes for which the average sign is reasonable and its deviation from unity can be mitigated by longer simulation times and intermediate temperatures. In the main text we study the half-filling case; in the SM, we report the average signs and results for DQMC away from half-filling, along with information about the DQMC simulation.

\begin{figure}[!t]
\centerline{\includegraphics[width=0.43\textwidth]{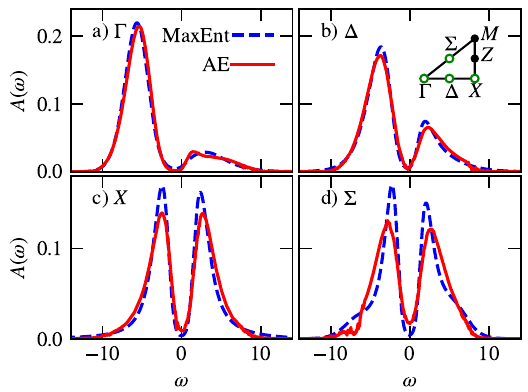}}
\caption{(a)-(d) DQMC related SFs calculated with the AE (red lines) and MaxEnt (blue dashed lines) for different momenta at high symmetry points, marked on panel (b) for convenience. The exact momenta used are indicated with open green circles. These results are for $\mu=0$, $\beta = 2t$, and $U/t=8$. \label{fig:A_K}}
\end{figure}

\begin{figure}[!ht]
\centerline{\includegraphics[width=0.5\textwidth]{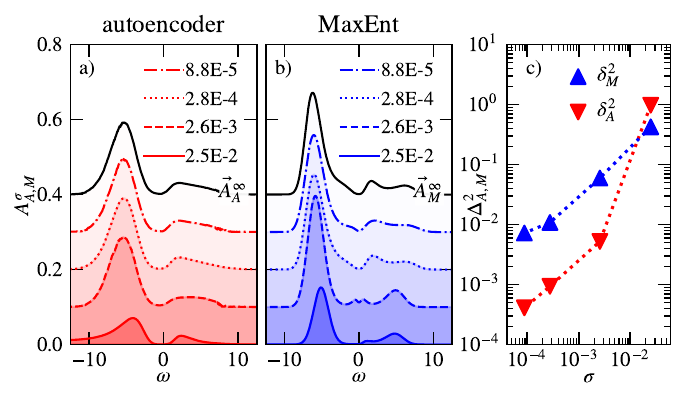}}
\caption{(a) and (b) Examples of SFs obtained by AE and MaxEnt. The SFs are overlaid, starting from the shortest QMC runs (highest $\sigma$). $A^{\infty}_{A,M}$ are indicated with black lines. (c) Convergence rates. The distance is defined in Eq.~\eqref{eq:deltas1}. The AE [MaxEnt] results are marked with red inverse triangles [blue triangles]. We average over the errors in $G(\tau)$'s at all $\tau$ to obtain $\sigma$. Without a specific selection, we show the SFs at the $\Gamma$ point within the Brillouin zone. The dotted lines in panel (c) are guides to the eye.\label{fig:conver}}
\end{figure}

We begin the discussion of DQMC results by showing SFs $A_{\vec{k}}(\omega)$ calculated on a $16\times 16$ system at high symmetry points of the Brillouin zone in Fig.~\ref{fig:A_K}. Both AE and MaxEnt results are presented, with relatively good agreement between these two approaches. Similarly, in the SM we show the comparison of the AE and MaxEnt density of states.
However, since in this case, the exact SF is not known, it is not possible to tell which of these results is more accurate. Therefore, it is important to propose another method to estimate the accuracy in the proposed approach. To this end, we compare the convergence rate of the AE with that of the MaxEnt method. In analogy to Eq.~\eqref{eq:deltas}, we define
\begin{equation}
\Delta_A^2(\sigma)=||\vec{A}_A^\infty-\vec{A}_A^\sigma||^2,\quad \Delta_M^2(\sigma)=||\vec{A}_M^\infty-\vec{A}_M^\sigma||^2,
\label{eq:deltas1}
\end{equation}
where $\vec{A}_A^\infty$ [$\vec{A}_M^\infty$] denotes the SF obtained within the AE [MaxEnt] method for the number of $G(\tau)$'s sufficiently large to ensure the SF's convergence. $\vec{A}_A^\sigma$ and $\vec{A}_M^\sigma$ represent SFs analogous to $\vec{A}_A^\infty$ and $\vec{A}_M^\infty$, but are calculated from a smaller number of $G(\tau)$'s, which is chosen to give a statistical error equal to $\sigma$. Figure~\ref{fig:conver} shows the convergence rates for AE and MaxEnt.
The AE SFs converge much faster than the MaxEnt SFs, i.e., the AE method can produce the same quality SFs from more noisy $G(\tau)$'s than the MaxEnt method. Thus, in practical applications, shorter QMC runs would be required when using the AE.
\vskip 2mm
\noindent \ul{\it Summary:}
The results presented demonstrate the following important features of the proposed autoencoder approach for analytic continuation:\\[1.0mm]
$\bullet$ In contrast to MaxEnt, where one has to define the default model, no prior knowledge about the physical nature
of the SFs is needed.\\[1.0mm]
$\bullet$ In the standard approach, MaxEnt is applied to smooth $G(\tau)$'s obtained by averaged QMC runs. In the AE method, the NN aims to recognize the meaningful characteristics of $G(\tau)$'s and distinguish them from statistical noise. More efficiently, unlike basic averaging, which necessitates numerous instances with different noise realizations for effective denoising, a properly trained AE can do with just a single image \cite{vincent_stacked_2010,buades_review_2005}. That is why, in most cases, for larger $\sigma$ the AE outperforms MaxEnt [cf. Figs.~\ref{fig:errors} (e) and (f)]. The ability to obtain the same precision in QMC through shorter simulations is crucial, especially concerning the fermionic sign problem.\\[1.0mm]
$\bullet$ By introducing custom regularization, imposing physical constraints on the SFs is straightforward. In addition to nonnegativity or smoothness requirements, one can impose, e.g., the sum rules~\cite{white_spectral_1991} (see SM). In Table~\ref{tab:sum_rules}, we compare the deviations of the AE's and MaxEnt's SFs three lowest moments from the exact values.
While the zeroth moment (normalization of the SF) is rather precise in both approaches, the errors for remaining moments are significantly smaller for AE.\\[1.0mm]
$\bullet$ The AE requires only one training for a given set of model parameters, such as $U$ or filling. Training can, in principle, be carried out for all momenta at the same time, which helps to avoid overfitting. Usually, the AE does not need to be re-trained for small changes in model. After the training stage, a few $G(\tau)$'s are sufficient to produce a reliable SF.
\begin{table}[!ht]
    \centering
    \begin{tabular}{|c||c|c|c|c|c|c|c|c|}\hline
        \multirow{2}{*}{$m$} 
               & \multicolumn{4}{c|}{AE}  & \multicolumn{4}{c|}{MaxEnt}\\
                \cline{2-9}
                & $\Gamma$ & $\Delta$ & $X$ &$\Sigma$ &$\Gamma$ & $\Delta$ & $X$ &$\Sigma$ \\
                \hline
        $0$ & $0.0073$ & 0.0053 & 0.0009 & 0.003 & 0.002 & 0.0014 & 0.002 & 0.0011 \\
        $1$ & $0.02$ & 0.0012 & 0.0001 & 0.0021 & 0.07 & 0.024 & 0.02 & 0.0055 \\
        $2$ & $0.61$ & 0.16 & 0.71 & 0.12 & 1.6 & 1.5 & 3.6 & 0.7 \\
        \hline
    \end{tabular}
    \caption{Absolute errors of $m^{\rm th}$ order moments $\mu^m_{\vec{k}} = \int_{-\infty}^{\infty}\omega ^m A_{\vec{k}}(\omega)  d\omega$ for various momenta $\vec{k}$, obtained with AE and MaxEnt from the QMC SFs. Their analytical formulas can be found in SM.}
    \label{tab:sum_rules}
\end{table}

In conclusion, we demonstrated the capabilities of the proposed method for extracting SFs from $G(\tau)$'s using AEs. It was done without making any prior assumptions about the SFs. We acknowledge the previous work and highlight that the generative framework may introduce more suitable initial conditions for the later training of the AE. By using the $G(\tau)$'s obtained from QMC only, we show the ability to grasp the spectral information from the imaginary time correlators. We compared the performance of AE with the MaxEnt method. We observed that AE equals or outperforms MaxEnt. The AE's and MaxEnt's predictions show similar features. However, our method exhibits higher robustness to statistical noise. Remarkably, as a result, AE can be applied to data from notably shorter simulations, enabling the study of larger systems. Our method offers valuable insights into the spectral properties of quantum systems, particularly when QMC simulations are computationally demanding.

This work highlights and upholds the potential of using AE as a powerful tool for extracting SFs from Green's functions, paving the way for advances in condensed matter physics and other fields where such spectral analysis is crucial. We anticipate that the advancement of more sophisticated networks within the rapidly evolving field of ML will incorporate and build on the methodology proposed in this study, leading to a deeper understanding of strongly correlated systems. We notice that numerous diverse ideas, recently introduced, can be integrated into the framework of this approach.

\begin{acknowledgments}
M.M.M. and M.K. acknowledge support from the National Science Centre (Poland) under Grant No. DEC-2018/29/B/ST3/01892. 
L.K., M.R., and N.T. were supported by the NSF Materials Research Science and Engineering Center Grant No. DMR-2011876. S.R. was supported by NSF grant no GR126818. T.P. acknowledges financial support from 
Funda\c{c}\~ao Carlos Chagas Filho de Amparo \`a Pesquisa do Estado do Rio de Janeiro grant numbers E-26/200.959/2022  and E-26/210.100/2023; CNPq grant numbers, 403130/2021-2  and 308335/2019-8; and also INCT-IQ. Numerical calculations have been partly carried out using high-performance computing resources provided by the Wrocław Centre for Networking and Supercomputing.
\end{acknowledgments}


\begin{thebibliography}{50}%
\makeatletter
\providecommand \@ifxundefined [1]{%
 \@ifx{#1\undefined}
}%
\providecommand \@ifnum [1]{%
 \ifnum #1\expandafter \@firstoftwo
 \else \expandafter \@secondoftwo
 \fi
}%
\providecommand \@ifx [1]{%
 \ifx #1\expandafter \@firstoftwo
 \else \expandafter \@secondoftwo
 \fi
}%
\providecommand \natexlab [1]{#1}%
\providecommand \enquote  [1]{``#1''}%
\providecommand \bibnamefont  [1]{#1}%
\providecommand \bibfnamefont [1]{#1}%
\providecommand \citenamefont [1]{#1}%
\providecommand \href@noop [0]{\@secondoftwo}%
\providecommand \href [0]{\begingroup \@sanitize@url \@href}%
\providecommand \@href[1]{\@@startlink{#1}\@@href}%
\providecommand \@@href[1]{\endgroup#1\@@endlink}%
\providecommand \@sanitize@url [0]{\catcode `\\12\catcode `\$12\catcode
  `\&12\catcode `\#12\catcode `\^12\catcode `\_12\catcode `\%12\relax}%
\providecommand \@@startlink[1]{}%
\providecommand \@@endlink[0]{}%
\providecommand \url  [0]{\begingroup\@sanitize@url \@url }%
\providecommand \@url [1]{\endgroup\@href {#1}{\urlprefix }}%
\providecommand \urlprefix  [0]{URL }%
\providecommand \Eprint [0]{\href }%
\providecommand \doibase [0]{https://doi.org/}%
\providecommand \selectlanguage [0]{\@gobble}%
\providecommand \bibinfo  [0]{\@secondoftwo}%
\providecommand \bibfield  [0]{\@secondoftwo}%
\providecommand \translation [1]{[#1]}%
\providecommand \BibitemOpen [0]{}%
\providecommand \bibitemStop [0]{}%
\providecommand \bibitemNoStop [0]{.\EOS\space}%
\providecommand \EOS [0]{\spacefactor3000\relax}%
\providecommand \BibitemShut  [1]{\csname bibitem#1\endcsname}%
\let\auto@bib@innerbib\@empty
\bibitem [{\citenamefont {Kabanikhin}(2011)}]{kabanikhin_inverse_2011}%
  \BibitemOpen
  \bibfield  {author} {\bibinfo {author} {\bibfnamefont {S.~I.}\ \bibnamefont
  {Kabanikhin}},\ }in\ \href
  {https://www.degruyter.com/document/doi/10.1515/9783110224016/html} {\emph
  {\bibinfo {booktitle} {Inverse and {Ill}-posed {Problems}}}}\ (\bibinfo
  {publisher} {De Gruyter},\ \bibinfo {year} {2011})\BibitemShut {NoStop}%
\bibitem [{\citenamefont {Tanaka}\ \emph {et~al.}(2021)\citenamefont {Tanaka},
  \citenamefont {Tomiya},\ and\ \citenamefont
  {Hashimoto}}]{tanaka_inverse_2021}%
  \BibitemOpen
  \bibfield  {author} {\bibinfo {author} {\bibfnamefont {A.}~\bibnamefont
  {Tanaka}}, \bibinfo {author} {\bibfnamefont {A.}~\bibnamefont {Tomiya}},\
  and\ \bibinfo {author} {\bibfnamefont {K.}~\bibnamefont {Hashimoto}},\ }in\
  \href {https://doi.org/10.1007/978-981-33-6108-9_7} {\emph {\bibinfo
  {booktitle} {Deep {Learning} and {Physics}}}},\ \bibinfo {series and number}
  {Mathematical {Physics} {Studies}},\ \bibinfo {editor} {edited by\ \bibinfo
  {editor} {\bibfnamefont {A.}~\bibnamefont {Tanaka}}, \bibinfo {editor}
  {\bibfnamefont {A.}~\bibnamefont {Tomiya}},\ and\ \bibinfo {editor}
  {\bibfnamefont {K.}~\bibnamefont {Hashimoto}}}\ (\bibinfo  {publisher}
  {Springer},\ \bibinfo {address} {Singapore},\ \bibinfo {year} {2021})\ pp.\
  \bibinfo {pages} {129--138}\BibitemShut {NoStop}%
\bibitem [{\citenamefont {Tarantola}(1987)}]{tarantola_inverse_1987}%
  \BibitemOpen
  \bibfield  {author} {\bibinfo {author} {\bibfnamefont {A.}~\bibnamefont
  {Tarantola}},\ }\href@noop {} {\emph {\bibinfo {title} {Inverse problem
  theory: methods for data fitting and model parameter estimation}}}\ (\bibinfo
   {publisher} {Elsevier ; Distributors for the United States and Canada,
  Elsevier Science Pub. Co},\ \bibinfo {address} {Amsterdam ; New York : New
  York, NY, U.S.A},\ \bibinfo {year} {1987})\BibitemShut {NoStop}%
\bibitem [{\citenamefont {Jacquelin}\ \emph {et~al.}(2003)\citenamefont
  {Jacquelin}, \citenamefont {Bennani},\ and\ \citenamefont
  {Hamelin}}]{jacquelin_force_2003}%
  \BibitemOpen
  \bibfield  {author} {\bibinfo {author} {\bibfnamefont {E.}~\bibnamefont
  {Jacquelin}}, \bibinfo {author} {\bibfnamefont {A.}~\bibnamefont {Bennani}},\
  and\ \bibinfo {author} {\bibfnamefont {P.}~\bibnamefont {Hamelin}},\ }\href
  {https://doi.org/10.1016/S0022-460X(02)01441-4} {\bibfield  {journal}
  {\bibinfo  {journal} {Journal of Sound and Vibration}\ }\textbf {\bibinfo
  {volume} {265}},\ \bibinfo {pages} {81} (\bibinfo {year} {2003})}\BibitemShut
  {NoStop}%
\bibitem [{\citenamefont {Asim}\ \emph {et~al.}(2019)\citenamefont {Asim},
  \citenamefont {Shamshad},\ and\ \citenamefont {Ahmed}}]{asim_blind_2019}%
  \BibitemOpen
  \bibfield  {author} {\bibinfo {author} {\bibfnamefont {M.}~\bibnamefont
  {Asim}}, \bibinfo {author} {\bibfnamefont {F.}~\bibnamefont {Shamshad}},\
  and\ \bibinfo {author} {\bibfnamefont {A.}~\bibnamefont {Ahmed}},\ }\href
  {http://arxiv.org/abs/1802.04073} {\bibinfo {title} {Blind {Image}
  {Deconvolution} using {Deep} {Generative} {Priors}}} (\bibinfo {year}
  {2019}),\ \bibinfo {note} {arXiv:1802.04073 [cs]}\BibitemShut {NoStop}%
\bibitem [{\citenamefont {Adke}\ \emph {et~al.}(2021)\citenamefont {Adke},
  \citenamefont {Karnik}, \citenamefont {Berman},\ and\ \citenamefont
  {Mathi}}]{adke_detection_2021}%
  \BibitemOpen
  \bibfield  {author} {\bibinfo {author} {\bibfnamefont {D.}~\bibnamefont
  {Adke}}, \bibinfo {author} {\bibfnamefont {A.}~\bibnamefont {Karnik}},
  \bibinfo {author} {\bibfnamefont {H.}~\bibnamefont {Berman}},\ and\ \bibinfo
  {author} {\bibfnamefont {S.}~\bibnamefont {Mathi}},\ }in\ \href
  {https://doi.org/10.1109/ICAICST53116.2021.9497841} {\emph {\bibinfo
  {booktitle} {2021 {International} {Conference} on {Artificial} {Intelligence}
  and {Computer} {Science} {Technology} ({ICAICST})}}}\ (\bibinfo  {publisher}
  {IEEE},\ \bibinfo {address} {Yogyakarta, Indonesia},\ \bibinfo {year}
  {2021})\ pp.\ \bibinfo {pages} {79--83}\BibitemShut {NoStop}%
\bibitem [{\citenamefont {Hörmann}\ and\ \citenamefont
  {Jammoul}(2022)}]{hormann_consistently_2022}%
  \BibitemOpen
  \bibfield  {author} {\bibinfo {author} {\bibfnamefont {S.}~\bibnamefont
  {Hörmann}}\ and\ \bibinfo {author} {\bibfnamefont {F.}~\bibnamefont
  {Jammoul}},\ }\href {https://doi.org/10.1016/j.jmva.2021.104886} {\bibfield
  {journal} {\bibinfo  {journal} {Journal of Multivariate Analysis}\ }\textbf
  {\bibinfo {volume} {189}},\ \bibinfo {pages} {104886} (\bibinfo {year}
  {2022})}\BibitemShut {NoStop}%
\bibitem [{\citenamefont {De~Hoog}\ \emph {et~al.}(1982)\citenamefont
  {De~Hoog}, \citenamefont {Knight},\ and\ \citenamefont
  {Stokes}}]{de_hoog_improved_1982}%
  \BibitemOpen
  \bibfield  {author} {\bibinfo {author} {\bibfnamefont {F.~R.}\ \bibnamefont
  {De~Hoog}}, \bibinfo {author} {\bibfnamefont {J.~H.}\ \bibnamefont
  {Knight}},\ and\ \bibinfo {author} {\bibfnamefont {A.~N.}\ \bibnamefont
  {Stokes}},\ }\href {https://doi.org/10.1137/0903022} {\bibfield  {journal}
  {\bibinfo  {journal} {SIAM Journal on Scientific and Statistical Computing}\
  }\textbf {\bibinfo {volume} {3}},\ \bibinfo {pages} {357} (\bibinfo {year}
  {1982})}\BibitemShut {NoStop}%
\bibitem [{\citenamefont {Van~Iseghem}(1987)}]{van_iseghem_laplace_1987}%
  \BibitemOpen
  \bibfield  {author} {\bibinfo {author} {\bibfnamefont {J.}~\bibnamefont
  {Van~Iseghem}},\ }\href {https://doi.org/10.1016/0168-9274(87)90046-8}
  {\bibfield  {journal} {\bibinfo  {journal} {Applied Numerical Mathematics}\
  }\textbf {\bibinfo {volume} {3}},\ \bibinfo {pages} {529} (\bibinfo {year}
  {1987})}\BibitemShut {NoStop}%
\bibitem [{\citenamefont {Kwok}\ and\ \citenamefont
  {Barthez}(1989)}]{kwok_algorithm_1989}%
  \BibitemOpen
  \bibfield  {author} {\bibinfo {author} {\bibfnamefont {Y.-K.}\ \bibnamefont
  {Kwok}}\ and\ \bibinfo {author} {\bibfnamefont {D.}~\bibnamefont {Barthez}},\
  }\href {https://doi.org/10.1088/0266-5611/5/6/014} {\bibfield  {journal}
  {\bibinfo  {journal} {Inverse Problems}\ }\textbf {\bibinfo {volume} {5}},\
  \bibinfo {pages} {1089} (\bibinfo {year} {1989})}\BibitemShut {NoStop}%
\bibitem [{\citenamefont {Gzyl}(1995)}]{gzyl_maxentropic_1995}%
  \BibitemOpen
  \bibfield  {author} {\bibinfo {author} {\bibfnamefont {H.}~\bibnamefont
  {Gzyl}},\ }\href {https://doi.org/10.1016/0096-3003(94)00248-7} {\bibfield
  {journal} {\bibinfo  {journal} {Applied Mathematics and Computation}\
  }\textbf {\bibinfo {volume} {73}},\ \bibinfo {pages} {181} (\bibinfo {year}
  {1995})}\BibitemShut {NoStop}%
\bibitem [{\citenamefont {Scotti}\ and\ \citenamefont
  {Wirgin}(1995)}]{scotti_shape_1995}%
  \BibitemOpen
  \bibfield  {author} {\bibinfo {author} {\bibfnamefont {T.}~\bibnamefont
  {Scotti}}\ and\ \bibinfo {author} {\bibfnamefont {A.}~\bibnamefont
  {Wirgin}},\ }\href {https://doi.org/10.1088/0266-5611/11/5/013} {\bibfield
  {journal} {\bibinfo  {journal} {Inverse Problems}\ }\textbf {\bibinfo
  {volume} {11}},\ \bibinfo {pages} {1097} (\bibinfo {year}
  {1995})}\BibitemShut {NoStop}%
\bibitem [{\citenamefont {Scotti}\ and\ \citenamefont
  {Wirgin}(1996)}]{scotti_shape_1996}%
  \BibitemOpen
  \bibfield  {author} {\bibinfo {author} {\bibfnamefont {T.}~\bibnamefont
  {Scotti}}\ and\ \bibinfo {author} {\bibfnamefont {A.}~\bibnamefont
  {Wirgin}},\ }\href {https://doi.org/10.1088/0266-5611/12/6/015} {\bibfield
  {journal} {\bibinfo  {journal} {Inverse Problems}\ }\textbf {\bibinfo
  {volume} {12}},\ \bibinfo {pages} {1027} (\bibinfo {year}
  {1996})}\BibitemShut {NoStop}%
\bibitem [{\citenamefont {Kant}\ and\ \citenamefont {Rudolf
  Von~Rohr}(2016)}]{kant_determination_2016}%
  \BibitemOpen
  \bibfield  {author} {\bibinfo {author} {\bibfnamefont {M.~A.}\ \bibnamefont
  {Kant}}\ and\ \bibinfo {author} {\bibfnamefont {P.}~\bibnamefont {Rudolf
  Von~Rohr}},\ }\href
  {https://doi.org/10.1016/j.ijheatmasstransfer.2016.03.082} {\bibfield
  {journal} {\bibinfo  {journal} {International Journal of Heat and Mass
  Transfer}\ }\textbf {\bibinfo {volume} {99}},\ \bibinfo {pages} {1} (\bibinfo
  {year} {2016})}\BibitemShut {NoStop}%
\bibitem [{\citenamefont {Lucy}(1974)}]{lucy_iterative_1974}%
  \BibitemOpen
  \bibfield  {author} {\bibinfo {author} {\bibfnamefont {L.~B.}\ \bibnamefont
  {Lucy}},\ }\href {https://doi.org/10.1086/111605} {\bibfield  {journal}
  {\bibinfo  {journal} {The Astronomical Journal}\ }\textbf {\bibinfo {volume}
  {79}},\ \bibinfo {pages} {745} (\bibinfo {year} {1974})}\BibitemShut
  {NoStop}%
\bibitem [{\citenamefont {Richardson}(1972)}]{richardson_bayesian-based_1972}%
  \BibitemOpen
  \bibfield  {author} {\bibinfo {author} {\bibfnamefont {W.~H.}\ \bibnamefont
  {Richardson}},\ }\href {https://doi.org/10.1364/JOSA.62.000055} {\bibfield
  {journal} {\bibinfo  {journal} {JOSA}\ }\textbf {\bibinfo {volume} {62}},\
  \bibinfo {pages} {55} (\bibinfo {year} {1972})}\BibitemShut {NoStop}%
\bibitem [{\citenamefont {Polianin}\ and\ \citenamefont
  {Manzhirov}(1998)}]{polianin_handbook_1998}%
  \BibitemOpen
  \bibfield  {author} {\bibinfo {author} {\bibfnamefont {A.~D.}\ \bibnamefont
  {Polianin}}\ and\ \bibinfo {author} {\bibfnamefont {A.~V.}\ \bibnamefont
  {Manzhirov}},\ }\href@noop {} {\emph {\bibinfo {title} {Handbook of integral
  equations}}}\ (\bibinfo  {publisher} {CRC Press},\ \bibinfo {address} {Boca
  Raton, Fla},\ \bibinfo {year} {1998})\BibitemShut {NoStop}%
\bibitem [{\citenamefont {Zappala}\ \emph {et~al.}(2023)\citenamefont
  {Zappala}, \citenamefont {Fonseca}, \citenamefont {Caro},\ and\ \citenamefont
  {van Dijk}}]{zappala_neural_2023}%
  \BibitemOpen
  \bibfield  {author} {\bibinfo {author} {\bibfnamefont {E.}~\bibnamefont
  {Zappala}}, \bibinfo {author} {\bibfnamefont {A.~H. d.~O.}\ \bibnamefont
  {Fonseca}}, \bibinfo {author} {\bibfnamefont {J.~O.}\ \bibnamefont {Caro}},\
  and\ \bibinfo {author} {\bibfnamefont {D.}~\bibnamefont {van Dijk}},\ }\href
  {http://arxiv.org/abs/2209.15190} {\bibinfo {title} {Neural {Integral}
  {Equations}}} (\bibinfo {year} {2023}),\ \bibinfo {note} {arXiv:2209.15190
  [physics]}\BibitemShut {NoStop}%
\bibitem [{\citenamefont {Guan}\ \emph {et~al.}(2022)\citenamefont {Guan},
  \citenamefont {Fang}, \citenamefont {Zhang},\ and\ \citenamefont
  {Jin}}]{guan_solving_2022}%
  \BibitemOpen
  \bibfield  {author} {\bibinfo {author} {\bibfnamefont {Y.}~\bibnamefont
  {Guan}}, \bibinfo {author} {\bibfnamefont {T.}~\bibnamefont {Fang}}, \bibinfo
  {author} {\bibfnamefont {D.}~\bibnamefont {Zhang}},\ and\ \bibinfo {author}
  {\bibfnamefont {C.}~\bibnamefont {Jin}},\ }\href
  {https://doi.org/10.1007/s40819-022-01288-3} {\bibfield  {journal} {\bibinfo
  {journal} {International Journal of Applied and Computational Mathematics}\
  }\textbf {\bibinfo {volume} {8}},\ \bibinfo {pages} {87} (\bibinfo {year}
  {2022})}\BibitemShut {NoStop}%
\bibitem [{\citenamefont {Zhang}\ \emph {et~al.}(2022)\citenamefont {Zhang},
  \citenamefont {Merkel}, \citenamefont {Beck},\ and\ \citenamefont
  {Ederer}}]{zhang_training_2022}%
  \BibitemOpen
  \bibfield  {author} {\bibinfo {author} {\bibfnamefont {R.}~\bibnamefont
  {Zhang}}, \bibinfo {author} {\bibfnamefont {M.~E.}\ \bibnamefont {Merkel}},
  \bibinfo {author} {\bibfnamefont {S.}~\bibnamefont {Beck}},\ and\ \bibinfo
  {author} {\bibfnamefont {C.}~\bibnamefont {Ederer}},\ }\href
  {https://doi.org/10.1103/PhysRevResearch.4.043082} {\bibfield  {journal}
  {\bibinfo  {journal} {Physical Review Research}\ }\textbf {\bibinfo {volume}
  {4}},\ \bibinfo {pages} {043082} (\bibinfo {year} {2022})}\BibitemShut
  {NoStop}%
\bibitem [{\citenamefont {Shao}\ and\ \citenamefont
  {Sandvik}(2023)}]{shao_progress_2023}%
  \BibitemOpen
  \bibfield  {author} {\bibinfo {author} {\bibfnamefont {H.}~\bibnamefont
  {Shao}}\ and\ \bibinfo {author} {\bibfnamefont {A.~W.}\ \bibnamefont
  {Sandvik}},\ }\href {https://doi.org/10.1016/j.physrep.2022.11.002}
  {\bibfield  {journal} {\bibinfo  {journal} {Physics Reports}\ }\textbf
  {\bibinfo {volume} {1003}},\ \bibinfo {pages} {1} (\bibinfo {year}
  {2023})}\BibitemShut {NoStop}%
\bibitem [{\citenamefont {Rao}\ and\ \citenamefont
  {Mitra}(1971)}]{rao_generalized_1971}%
  \BibitemOpen
  \bibfield  {author} {\bibinfo {author} {\bibfnamefont {C.~R.}\ \bibnamefont
  {Rao}}\ and\ \bibinfo {author} {\bibfnamefont {S.~K.}\ \bibnamefont
  {Mitra}},\ }\href@noop {} {\emph {\bibinfo {title} {Generalized inverse of
  matrices and its applications}}},\ Wiley series in probability and
  mathematical statistics\ (\bibinfo  {publisher} {Wiley},\ \bibinfo {address}
  {New York},\ \bibinfo {year} {1971})\BibitemShut {NoStop}%
\bibitem [{\citenamefont {Ben-Israel}\ and\ \citenamefont
  {Greville}(2003)}]{ben-israel_generalized_2003}%
  \BibitemOpen
  \bibfield  {author} {\bibinfo {author} {\bibfnamefont {A.}~\bibnamefont
  {Ben-Israel}}\ and\ \bibinfo {author} {\bibfnamefont {T.~N.~E.}\ \bibnamefont
  {Greville}},\ }\href@noop {} {{\selectlanguage {eng}\emph {\bibinfo {title}
  {Generalized inverses: theory and applications}}}},\ \bibinfo {edition}
  {2nd}\ ed.,\ \bibinfo {series} {{CMS} books in mathematics}\ No.~\bibinfo
  {number} {15}\ (\bibinfo  {publisher} {Springer},\ \bibinfo {address} {New
  York, NY},\ \bibinfo {year} {2003})\BibitemShut {NoStop}%
\bibitem [{\citenamefont {Schott}\ \emph {et~al.}(2016)\citenamefont {Schott},
  \citenamefont {Locht}, \citenamefont {Lundin}, \citenamefont {Granas},
  \citenamefont {Eriksson},\ and\ \citenamefont
  {Di~Marco}}]{schott_analytic_2016}%
  \BibitemOpen
  \bibfield  {author} {\bibinfo {author} {\bibfnamefont {J.}~\bibnamefont
  {Schott}}, \bibinfo {author} {\bibfnamefont {I.~L.~M.}\ \bibnamefont
  {Locht}}, \bibinfo {author} {\bibfnamefont {E.}~\bibnamefont {Lundin}},
  \bibinfo {author} {\bibfnamefont {O.}~\bibnamefont {Granas}}, \bibinfo
  {author} {\bibfnamefont {O.}~\bibnamefont {Eriksson}},\ and\ \bibinfo
  {author} {\bibfnamefont {I.}~\bibnamefont {Di~Marco}},\ }\href
  {https://doi.org/10.1103/PhysRevB.93.075104} {\bibfield  {journal} {\bibinfo
  {journal} {Physical Review B}\ }\textbf {\bibinfo {volume} {93}},\ \bibinfo
  {pages} {075104} (\bibinfo {year} {2016})}\BibitemShut {NoStop}%
\bibitem [{\citenamefont {Sandvik}(1998)}]{sandvik_stochastic_1998}%
  \BibitemOpen
  \bibfield  {author} {\bibinfo {author} {\bibfnamefont {A.~W.}\ \bibnamefont
  {Sandvik}},\ }\href {https://doi.org/10.1103/PhysRevB.57.10287} {\bibfield
  {journal} {\bibinfo  {journal} {Physical Review B}\ }\textbf {\bibinfo
  {volume} {57}},\ \bibinfo {pages} {10287} (\bibinfo {year}
  {1998})}\BibitemShut {NoStop}%
\bibitem [{\citenamefont {Otsuki}\ \emph {et~al.}(2017)\citenamefont {Otsuki},
  \citenamefont {Ohzeki}, \citenamefont {Shinaoka},\ and\ \citenamefont
  {Yoshimi}}]{otsuki_sparse_2017}%
  \BibitemOpen
  \bibfield  {author} {\bibinfo {author} {\bibfnamefont {J.}~\bibnamefont
  {Otsuki}}, \bibinfo {author} {\bibfnamefont {M.}~\bibnamefont {Ohzeki}},
  \bibinfo {author} {\bibfnamefont {H.}~\bibnamefont {Shinaoka}},\ and\
  \bibinfo {author} {\bibfnamefont {K.}~\bibnamefont {Yoshimi}},\ }\href
  {https://doi.org/10.1103/PhysRevE.95.061302} {\bibfield  {journal} {\bibinfo
  {journal} {Physical Review E}\ }\textbf {\bibinfo {volume} {95}},\ \bibinfo
  {pages} {061302(R)} (\bibinfo {year} {2017})}\BibitemShut {NoStop}%
\bibitem [{\citenamefont {Yoshimi}(2019)}]{yoshimi_spm_2019}%
  \BibitemOpen
  \bibfield  {author} {\bibinfo {author} {\bibfnamefont {K.}~\bibnamefont
  {Yoshimi}},\ }\href {https://doi.org/10.17632/YCMPSNV5YX.1} {\bibinfo {title}
  {{SpM}: {Sparse} modeling tool for analytic continuation of imaginary-time
  {Green}'s function}} (\bibinfo {year} {2019})\BibitemShut {NoStop}%
\bibitem [{\citenamefont {Motoyama}\ \emph {et~al.}(2022)\citenamefont
  {Motoyama}, \citenamefont {Yoshimi},\ and\ \citenamefont
  {Otsuki}}]{motoyama_robust_2022}%
  \BibitemOpen
  \bibfield  {author} {\bibinfo {author} {\bibfnamefont {Y.}~\bibnamefont
  {Motoyama}}, \bibinfo {author} {\bibfnamefont {K.}~\bibnamefont {Yoshimi}},\
  and\ \bibinfo {author} {\bibfnamefont {J.}~\bibnamefont {Otsuki}},\ }\href
  {https://doi.org/10.1103/PhysRevB.105.035139} {\bibfield  {journal} {\bibinfo
   {journal} {Physical Review B}\ }\textbf {\bibinfo {volume} {105}},\ \bibinfo
  {pages} {035139} (\bibinfo {year} {2022})}\BibitemShut {NoStop}%
\bibitem [{\citenamefont {White}(1991{\natexlab{a}})}]{lotsch_average_1991}%
  \BibitemOpen
  \bibfield  {author} {\bibinfo {author} {\bibfnamefont {S.~R.}\ \bibnamefont
  {White}},\ }in\ \href {https://doi.org/10.1007/978-3-642-76382-3_13} {\emph
  {\bibinfo {booktitle} {Computer {Simulation} {Studies} in {Condensed}
  {Matter} {Physics} {III}}}},\ Vol.~\bibinfo {volume} {53},\ \bibinfo {editor}
  {edited by\ \bibinfo {editor} {\bibfnamefont {H.~K.~V.}\ \bibnamefont
  {Lotsch}}, \bibinfo {editor} {\bibfnamefont {D.~P.}\ \bibnamefont {Landau}},
  \bibinfo {editor} {\bibfnamefont {K.~K.}\ \bibnamefont {Mon}},\ and\ \bibinfo
  {editor} {\bibfnamefont {H.-B.}\ \bibnamefont {Schüttler}}}\ (\bibinfo
  {publisher} {Springer Berlin Heidelberg},\ \bibinfo {address} {Berlin,
  Heidelberg},\ \bibinfo {year} {1991})\ pp.\ \bibinfo {pages} {145--153},\
  \bibinfo {note} {series Title: Springer Proceedings in Physics}\BibitemShut
  {NoStop}%
\bibitem [{\citenamefont {Huang}\ and\ \citenamefont
  {Liang}(2023)}]{huang_stochastic_2023}%
  \BibitemOpen
  \bibfield  {author} {\bibinfo {author} {\bibfnamefont {L.}~\bibnamefont
  {Huang}}\ and\ \bibinfo {author} {\bibfnamefont {S.}~\bibnamefont {Liang}},\
  }\href {https://doi.org/10.48550/arXiv.2307.11324} {\bibinfo {title}
  {Stochastic pole expansion method}} (\bibinfo {year} {2023}),\ \bibinfo
  {note} {arXiv:2307.11324 [cond-mat, physics:physics]}\BibitemShut {NoStop}%
\bibitem [{\citenamefont {Vitali}\ \emph {et~al.}(2010)\citenamefont {Vitali},
  \citenamefont {Rossi}, \citenamefont {Reatto},\ and\ \citenamefont
  {Galli}}]{vitali_ab_2010}%
  \BibitemOpen
  \bibfield  {author} {\bibinfo {author} {\bibfnamefont {E.}~\bibnamefont
  {Vitali}}, \bibinfo {author} {\bibfnamefont {M.}~\bibnamefont {Rossi}},
  \bibinfo {author} {\bibfnamefont {L.}~\bibnamefont {Reatto}},\ and\ \bibinfo
  {author} {\bibfnamefont {D.~E.}\ \bibnamefont {Galli}},\ }\href
  {https://doi.org/10.1103/PhysRevB.82.174510} {\bibfield  {journal} {\bibinfo
  {journal} {Physical Review B}\ }\textbf {\bibinfo {volume} {82}},\ \bibinfo
  {pages} {174510} (\bibinfo {year} {2010})}\BibitemShut {NoStop}%
\bibitem [{\citenamefont {Pavarini}\ \emph {et~al.}(2012)\citenamefont
  {Pavarini}, \citenamefont {Koch}, \citenamefont {Anders}, \citenamefont
  {Jarrell}, \citenamefont {for Advanced~Simulation},\ and\ \citenamefont {for
  Simulation~Sciences}}]{pavarini_correlated_2012}%
  \BibitemOpen
  \bibinfo {editor} {\bibfnamefont {E.}~\bibnamefont {Pavarini}}, \bibinfo
  {editor} {\bibfnamefont {E.}~\bibnamefont {Koch}}, \bibinfo {editor}
  {\bibfnamefont {F.}~\bibnamefont {Anders}}, \bibinfo {editor} {\bibfnamefont
  {M.}~\bibnamefont {Jarrell}}, \bibinfo {editor} {\bibfnamefont
  {I.}~\bibnamefont {for Advanced~Simulation}},\ and\ \bibinfo {editor}
  {\bibfnamefont {G.~R.~S.}\ \bibnamefont {for Simulation~Sciences}},\ eds.,\
  \href@noop {} {\emph {\bibinfo {title} {Correlated electrons: from models to
  materials: lecture notes of the {Autumn} {School} {Correlated} {Electrons}
  2012: at {Forschungszentrum} {Jülich}, 3-7 {September} 2012}}},\ \bibinfo
  {series} {Schriften des {Forschungszentrums} {Jülich}. {Reihe} {Modeling}
  and {Simulation}}\ No.\ \bibinfo {number} {Band 2}\ (\bibinfo  {publisher}
  {Forschungszentrum Jülich, Zentralbibliothek, Verl},\ \bibinfo {address}
  {Jülich},\ \bibinfo {year} {2012})\BibitemShut {NoStop}%
\bibitem [{\citenamefont {Levy}\ \emph {et~al.}(2017)\citenamefont {Levy},
  \citenamefont {LeBlanc},\ and\ \citenamefont
  {Gull}}]{levy_implementation_2017}%
  \BibitemOpen
  \bibfield  {author} {\bibinfo {author} {\bibfnamefont {R.}~\bibnamefont
  {Levy}}, \bibinfo {author} {\bibfnamefont {J.~P.~F.}\ \bibnamefont
  {LeBlanc}},\ and\ \bibinfo {author} {\bibfnamefont {E.}~\bibnamefont
  {Gull}},\ }\href {https://doi.org/10.17632/RF3P4PSDHS.1} {\bibinfo {title}
  {Implementation of the maximum entropy method for analytic continuation}}
  (\bibinfo {year} {2017}),\ \bibinfo {note}
  {10.17632/RF3P4PSDHS.1}\BibitemShut {NoStop}%
\bibitem [{\citenamefont {Fei}\ \emph {et~al.}(2021)\citenamefont {Fei},
  \citenamefont {Yeh},\ and\ \citenamefont {Gull}}]{fei_nevanlinna_2021}%
  \BibitemOpen
  \bibfield  {author} {\bibinfo {author} {\bibfnamefont {J.}~\bibnamefont
  {Fei}}, \bibinfo {author} {\bibfnamefont {C.-N.}\ \bibnamefont {Yeh}},\ and\
  \bibinfo {author} {\bibfnamefont {E.}~\bibnamefont {Gull}},\ }\href
  {https://doi.org/10.1103/PhysRevLett.126.056402} {\bibfield  {journal}
  {\bibinfo  {journal} {Physical Review Letters}\ }\textbf {\bibinfo {volume}
  {126}},\ \bibinfo {pages} {056402} (\bibinfo {year} {2021})}\BibitemShut
  {NoStop}%
\bibitem [{\citenamefont {Huang}\ \emph {et~al.}(2023)\citenamefont {Huang},
  \citenamefont {Gull},\ and\ \citenamefont {Lin}}]{huang_robust_2023}%
  \BibitemOpen
  \bibfield  {author} {\bibinfo {author} {\bibfnamefont {Z.}~\bibnamefont
  {Huang}}, \bibinfo {author} {\bibfnamefont {E.}~\bibnamefont {Gull}},\ and\
  \bibinfo {author} {\bibfnamefont {L.}~\bibnamefont {Lin}},\ }\href
  {https://doi.org/10.1103/PhysRevB.107.075151} {\bibfield  {journal} {\bibinfo
   {journal} {Physical Review B}\ }\textbf {\bibinfo {volume} {107}},\ \bibinfo
  {pages} {075151} (\bibinfo {year} {2023})}\BibitemShut {NoStop}%
\bibitem [{\citenamefont {Fournier}\ \emph {et~al.}(2020)\citenamefont
  {Fournier}, \citenamefont {Wang}, \citenamefont {Yazyev},\ and\ \citenamefont
  {Wu}}]{fournier_artificial_2020}%
  \BibitemOpen
  \bibfield  {author} {\bibinfo {author} {\bibfnamefont {R.}~\bibnamefont
  {Fournier}}, \bibinfo {author} {\bibfnamefont {L.}~\bibnamefont {Wang}},
  \bibinfo {author} {\bibfnamefont {O.~V.}\ \bibnamefont {Yazyev}},\ and\
  \bibinfo {author} {\bibfnamefont {Q.~S.}\ \bibnamefont {Wu}},\ }\href
  {https://doi.org/10.1103/PhysRevLett.124.056401} {\bibfield  {journal}
  {\bibinfo  {journal} {Physical Review Letters}\ }\textbf {\bibinfo {volume}
  {124}},\ \bibinfo {pages} {056401} (\bibinfo {year} {2020})}\BibitemShut
  {NoStop}%
\bibitem [{\citenamefont {Yoon}\ \emph {et~al.}(2018)\citenamefont {Yoon},
  \citenamefont {Sim},\ and\ \citenamefont {Han}}]{yoon_analytic_2018}%
  \BibitemOpen
  \bibfield  {author} {\bibinfo {author} {\bibfnamefont {H.}~\bibnamefont
  {Yoon}}, \bibinfo {author} {\bibfnamefont {J.-H.}\ \bibnamefont {Sim}},\ and\
  \bibinfo {author} {\bibfnamefont {M.~J.}\ \bibnamefont {Han}},\ }\href
  {https://doi.org/10.1103/PhysRevB.98.245101} {\bibfield  {journal} {\bibinfo
  {journal} {Physical Review B}\ }\textbf {\bibinfo {volume} {98}},\ \bibinfo
  {pages} {245101} (\bibinfo {year} {2018})}\BibitemShut {NoStop}%
\bibitem [{\citenamefont {Yao}\ \emph {et~al.}(2022)\citenamefont {Yao},
  \citenamefont {Wang}, \citenamefont {Yao},\ and\ \citenamefont
  {Zhai}}]{yao_noise_2022}%
  \BibitemOpen
  \bibfield  {author} {\bibinfo {author} {\bibfnamefont {J.}~\bibnamefont
  {Yao}}, \bibinfo {author} {\bibfnamefont {C.}~\bibnamefont {Wang}}, \bibinfo
  {author} {\bibfnamefont {Z.}~\bibnamefont {Yao}},\ and\ \bibinfo {author}
  {\bibfnamefont {H.}~\bibnamefont {Zhai}},\ }\href
  {https://doi.org/10.1088/2632-2153/ac6f44} {\bibfield  {journal} {\bibinfo
  {journal} {Machine Learning: Science and Technology}\ }\textbf {\bibinfo
  {volume} {3}},\ \bibinfo {pages} {025010} (\bibinfo {year}
  {2022})}\BibitemShut {NoStop}%
\bibitem [{\citenamefont {Hornik}\ \emph {et~al.}(1989)\citenamefont {Hornik},
  \citenamefont {Stinchcombe},\ and\ \citenamefont
  {White}}]{hornik_multilayer_1989}%
  \BibitemOpen
  \bibfield  {author} {\bibinfo {author} {\bibfnamefont {K.}~\bibnamefont
  {Hornik}}, \bibinfo {author} {\bibfnamefont {M.}~\bibnamefont
  {Stinchcombe}},\ and\ \bibinfo {author} {\bibfnamefont {H.}~\bibnamefont
  {White}},\ }\href {https://doi.org/10.1016/0893-6080(89)90020-8} {\bibfield
  {journal} {\bibinfo  {journal} {Neural Networks}\ }\textbf {\bibinfo {volume}
  {2}},\ \bibinfo {pages} {359} (\bibinfo {year} {1989})}\BibitemShut {NoStop}%
\bibitem [{\citenamefont {Siegelmann}\ and\ \citenamefont
  {Sontag}(1992)}]{siegelmann_computational_1992}%
  \BibitemOpen
  \bibfield  {author} {\bibinfo {author} {\bibfnamefont {H.~T.}\ \bibnamefont
  {Siegelmann}}\ and\ \bibinfo {author} {\bibfnamefont {E.~D.}\ \bibnamefont
  {Sontag}},\ }in\ \href {https://doi.org/10.1145/130385.130432}
  {{\selectlanguage {en}\emph {\bibinfo {booktitle} {Proceedings of the fifth
  annual workshop on {Computational} learning theory}}}}\ (\bibinfo
  {publisher} {ACM},\ \bibinfo {address} {Pittsburgh Pennsylvania USA},\
  \bibinfo {year} {1992})\ pp.\ \bibinfo {pages} {440--449}\BibitemShut
  {NoStop}%
\bibitem [{\citenamefont {Barron}(1993)}]{barron_universal_1993}%
  \BibitemOpen
  \bibfield  {author} {\bibinfo {author} {\bibfnamefont {A.}~\bibnamefont
  {Barron}},\ }\href {https://doi.org/10.1109/18.256500} {\bibfield  {journal}
  {\bibinfo  {journal} {IEEE Transactions on Information Theory}\ }\textbf
  {\bibinfo {volume} {39}},\ \bibinfo {pages} {930} (\bibinfo {year}
  {1993})}\BibitemShut {NoStop}%
\bibitem [{act()}]{actual_footnote}%
  \BibitemOpen
  \href@noop {} {\bibinfo {title} {In the actual calculations we use the
  covariance matrix to take into account different uncertainties of different
  {Green}'s functions.}}\BibitemShut {Stop}%
\bibitem [{sin()}]{since_footnote}%
  \BibitemOpen
  \href@noop {} {\bibinfo {title} {Since there are no uncertainties in the
  artificially generated {SFs}, contrary to minimizing the loss function
  defined in {Eq}. (5), here we do not use a covariance matrix.}}\BibitemShut
  {Stop}%
\bibitem [{\citenamefont {Bellman}\ and\ \citenamefont
  {Bellman}(1957)}]{bellman_dynamic_1957}%
  \BibitemOpen
  \bibfield  {author} {\bibinfo {author} {\bibfnamefont {R.}~\bibnamefont
  {Bellman}}\ and\ \bibinfo {author} {\bibfnamefont {R.}~\bibnamefont
  {Bellman}},\ }\href {https://books.google.pl/books?id=rZW4ugAACAAJ} {\emph
  {\bibinfo {title} {Dynamic programming}}},\ Rand corporation research study\
  (\bibinfo  {publisher} {Princeton University Press},\ \bibinfo {year}
  {1957})\BibitemShut {NoStop}%
\bibitem [{\citenamefont {White}(1991{\natexlab{b}})}]{white_spectral_1991}%
  \BibitemOpen
  \bibfield  {author} {\bibinfo {author} {\bibfnamefont {S.~R.}\ \bibnamefont
  {White}},\ }\href {https://doi.org/10.1103/PhysRevB.44.4670} {\bibfield
  {journal} {\bibinfo  {journal} {Physical Review B}\ }\textbf {\bibinfo
  {volume} {44}},\ \bibinfo {pages} {4670} (\bibinfo {year}
  {1991}{\natexlab{b}})}\BibitemShut {NoStop}%
\bibitem [{\citenamefont {Vincent}\ \emph {et~al.}(2010)\citenamefont
  {Vincent}, \citenamefont {Larochelle}, \citenamefont {Lajoie}, \citenamefont
  {Bengio},\ and\ \citenamefont {Manzagol}}]{vincent_stacked_2010}%
  \BibitemOpen
  \bibfield  {author} {\bibinfo {author} {\bibfnamefont {P.}~\bibnamefont
  {Vincent}}, \bibinfo {author} {\bibfnamefont {H.}~\bibnamefont {Larochelle}},
  \bibinfo {author} {\bibfnamefont {I.}~\bibnamefont {Lajoie}}, \bibinfo
  {author} {\bibfnamefont {Y.}~\bibnamefont {Bengio}},\ and\ \bibinfo {author}
  {\bibfnamefont {P.-A.}\ \bibnamefont {Manzagol}},\ }\href
  {http://jmlr.org/papers/v11/vincent10a.html} {\bibfield  {journal} {\bibinfo
  {journal} {Journal of Machine Learning Research}\ }\textbf {\bibinfo {volume}
  {11}},\ \bibinfo {pages} {3371} (\bibinfo {year} {2010})}\BibitemShut
  {NoStop}%
\bibitem [{\citenamefont {Troyer}\ and\ \citenamefont
  {Wiese}(2005)}]{troyer_computational_2005}%
  \BibitemOpen
  \bibfield  {author} {\bibinfo {author} {\bibfnamefont {M.}~\bibnamefont
  {Troyer}}\ and\ \bibinfo {author} {\bibfnamefont {U.-J.}\ \bibnamefont
  {Wiese}},\ }\href {https://doi.org/10.1103/PhysRevLett.94.170201} {\bibfield
  {journal} {\bibinfo  {journal} {Physical Review Letters}\ }\textbf {\bibinfo
  {volume} {94}},\ \bibinfo {pages} {170201} (\bibinfo {year}
  {2005})}\BibitemShut {NoStop}%
\bibitem [{\citenamefont {Hangleiter}\ \emph {et~al.}(2020)\citenamefont
  {Hangleiter}, \citenamefont {Roth}, \citenamefont {Nagaj},\ and\
  \citenamefont {Eisert}}]{hangleiter_easing_2020}%
  \BibitemOpen
  \bibfield  {author} {\bibinfo {author} {\bibfnamefont {D.}~\bibnamefont
  {Hangleiter}}, \bibinfo {author} {\bibfnamefont {I.}~\bibnamefont {Roth}},
  \bibinfo {author} {\bibfnamefont {D.}~\bibnamefont {Nagaj}},\ and\ \bibinfo
  {author} {\bibfnamefont {J.}~\bibnamefont {Eisert}},\ }\href
  {https://doi.org/10.1126/sciadv.abb8341} {\bibfield  {journal} {\bibinfo
  {journal} {Science Advances}\ }\textbf {\bibinfo {volume} {6}},\ \bibinfo
  {pages} {eabb8341} (\bibinfo {year} {2020})}\BibitemShut {NoStop}%
\bibitem [{\citenamefont {Pan}\ and\ \citenamefont
  {Meng}(2022)}]{pan_sign_2022}%
  \BibitemOpen
  \bibfield  {author} {\bibinfo {author} {\bibfnamefont {G.}~\bibnamefont
  {Pan}}\ and\ \bibinfo {author} {\bibfnamefont {Z.~Y.}\ \bibnamefont {Meng}},\
  }\href {http://arxiv.org/abs/2204.08777} {\bibinfo {title} {Sign {Problem} in
  {Quantum} {Monte} {Carlo} {Simulation}}} (\bibinfo {year} {2022}),\ \bibinfo
  {note} {arXiv:2204.08777 [cond-mat, physics:hep-lat]}\BibitemShut {NoStop}%
\bibitem [{\citenamefont {Buades}\ \emph {et~al.}(2005)\citenamefont {Buades},
  \citenamefont {Coll},\ and\ \citenamefont {Morel}}]{buades_review_2005}%
  \BibitemOpen
  \bibfield  {author} {\bibinfo {author} {\bibfnamefont {A.}~\bibnamefont
  {Buades}}, \bibinfo {author} {\bibfnamefont {B.}~\bibnamefont {Coll}},\ and\
  \bibinfo {author} {\bibfnamefont {J.~M.}\ \bibnamefont {Morel}},\ }\href
  {https://doi.org/10.1137/040616024} {\bibfield  {journal} {\bibinfo
  {journal} {Multiscale Modeling \& Simulation}\ }\textbf {\bibinfo {volume}
  {4}},\ \bibinfo {pages} {490} (\bibinfo {year} {2005})}\BibitemShut {NoStop}%
\end{thebibliography}%


\begin{thebibliography}{26}%
\makeatletter
\providecommand \@ifxundefined [1]{%
 \@ifx{#1\undefined}
}%
\providecommand \@ifnum [1]{%
 \ifnum #1\expandafter \@firstoftwo
 \else \expandafter \@secondoftwo
 \fi
}%
\providecommand \@ifx [1]{%
 \ifx #1\expandafter \@firstoftwo
 \else \expandafter \@secondoftwo
 \fi
}%
\providecommand \natexlab [1]{#1}%
\providecommand \enquote  [1]{``#1''}%
\providecommand \bibnamefont  [1]{#1}%
\providecommand \bibfnamefont [1]{#1}%
\providecommand \citenamefont [1]{#1}%
\providecommand \href@noop [0]{\@secondoftwo}%
\providecommand \href [0]{\begingroup \@sanitize@url \@href}%
\providecommand \@href[1]{\@@startlink{#1}\@@href}%
\providecommand \@@href[1]{\endgroup#1\@@endlink}%
\providecommand \@sanitize@url [0]{\catcode `\\12\catcode `\$12\catcode
  `\&12\catcode `\#12\catcode `\^12\catcode `\_12\catcode `\%12\relax}%
\providecommand \@@startlink[1]{}%
\providecommand \@@endlink[0]{}%
\providecommand \url  [0]{\begingroup\@sanitize@url \@url }%
\providecommand \@url [1]{\endgroup\@href {#1}{\urlprefix }}%
\providecommand \urlprefix  [0]{URL }%
\providecommand \Eprint [0]{\href }%
\providecommand \doibase [0]{https://doi.org/}%
\providecommand \selectlanguage [0]{\@gobble}%
\providecommand \bibinfo  [0]{\@secondoftwo}%
\providecommand \bibfield  [0]{\@secondoftwo}%
\providecommand \translation [1]{[#1]}%
\providecommand \BibitemOpen [0]{}%
\providecommand \bibitemStop [0]{}%
\providecommand \bibitemNoStop [0]{.\EOS\space}%
\providecommand \EOS [0]{\spacefactor3000\relax}%
\providecommand \BibitemShut  [1]{\csname bibitem#1\endcsname}%
\let\auto@bib@innerbib\@empty
\bibitem [{\citenamefont {Chollet}\ and\ \citenamefont
  {{others}}(2015)}]{chollet_keras_2015}%
  \BibitemOpen
  \bibfield  {author} {\bibinfo {author} {\bibfnamefont {F.}~\bibnamefont
  {Chollet}}\ and\ \bibinfo {author} {\bibnamefont {{others}}},\ }\href
  {https://keras.io/api/layers/} {\bibinfo {title} {Keras}} (\bibinfo {year}
  {2015})\BibitemShut {NoStop}%
\bibitem [{\citenamefont {{TensorFlow
  Developers}}(2023)}]{tensorflow_developers_tensorflow_2023}%
  \BibitemOpen
  \bibfield  {author} {\bibinfo {author} {\bibnamefont {{TensorFlow
  Developers}}},\ }\href {https://doi.org/10.5281/ZENODO.4724125} {\bibinfo
  {title} {{TensorFlow}}} (\bibinfo {year} {2023})\BibitemShut {NoStop}%
\bibitem [{\citenamefont {Zhang}\ \emph {et~al.}(2022)\citenamefont {Zhang},
  \citenamefont {Merkel}, \citenamefont {Beck},\ and\ \citenamefont
  {Ederer}}]{zhang_training_2022}%
  \BibitemOpen
  \bibfield  {author} {\bibinfo {author} {\bibfnamefont {R.}~\bibnamefont
  {Zhang}}, \bibinfo {author} {\bibfnamefont {M.~E.}\ \bibnamefont {Merkel}},
  \bibinfo {author} {\bibfnamefont {S.}~\bibnamefont {Beck}},\ and\ \bibinfo
  {author} {\bibfnamefont {C.}~\bibnamefont {Ederer}},\ }\href
  {https://doi.org/10.1103/PhysRevResearch.4.043082} {\bibfield  {journal}
  {\bibinfo  {journal} {Physical Review Research}\ }\textbf {\bibinfo {volume}
  {4}},\ \bibinfo {pages} {043082} (\bibinfo {year} {2022})}\BibitemShut
  {NoStop}%
\bibitem [{\citenamefont {Kingma}\ and\ \citenamefont
  {Welling}(2019)}]{kingma_introduction_2019}%
  \BibitemOpen
  \bibfield  {author} {\bibinfo {author} {\bibfnamefont {D.~P.}\ \bibnamefont
  {Kingma}}\ and\ \bibinfo {author} {\bibfnamefont {M.}~\bibnamefont
  {Welling}},\ }\href {https://doi.org/10.1561/2200000056} {\bibfield
  {journal} {\bibinfo  {journal} {Foundations and Trends® in Machine
  Learning}\ }\textbf {\bibinfo {volume} {12}},\ \bibinfo {pages} {307}
  (\bibinfo {year} {2019})}\BibitemShut {NoStop}%
\bibitem [{\citenamefont {Pinheiro~Cinelli}\ \emph {et~al.}(2021)\citenamefont
  {Pinheiro~Cinelli}, \citenamefont {Araújo~Marins}, \citenamefont {Barros
  Da~Silva},\ and\ \citenamefont {Lima~Netto}}]{cinelli_variational_2021}%
  \BibitemOpen
  \bibfield  {author} {\bibinfo {author} {\bibfnamefont {L.}~\bibnamefont
  {Pinheiro~Cinelli}}, \bibinfo {author} {\bibfnamefont {M.}~\bibnamefont
  {Araújo~Marins}}, \bibinfo {author} {\bibfnamefont {E.~A.}\ \bibnamefont
  {Barros Da~Silva}},\ and\ \bibinfo {author} {\bibfnamefont {S.}~\bibnamefont
  {Lima~Netto}},\ }in\ \href {https://doi.org/10.1007/978-3-030-70679-1_5}
  {\emph {\bibinfo {booktitle} {Variational {Methods} for {Machine} {Learning}
  with {Applications} to {Deep} {Networks}}}}\ (\bibinfo  {publisher} {Springer
  International Publishing},\ \bibinfo {address} {Cham},\ \bibinfo {year}
  {2021})\ pp.\ \bibinfo {pages} {111--149}\BibitemShut {NoStop}%
\bibitem [{\citenamefont {Xu}\ \emph {et~al.}(2018)\citenamefont {Xu},
  \citenamefont {Feng}, \citenamefont {Chen}, \citenamefont {Wang},
  \citenamefont {Qiao}, \citenamefont {Chen}, \citenamefont {Zhao},
  \citenamefont {Li}, \citenamefont {Bu}, \citenamefont {Li}, \citenamefont
  {Liu}, \citenamefont {Zhao},\ and\ \citenamefont
  {Pei}}]{xu_unsupervised_2018}%
  \BibitemOpen
  \bibfield  {author} {\bibinfo {author} {\bibfnamefont {H.}~\bibnamefont
  {Xu}}, \bibinfo {author} {\bibfnamefont {Y.}~\bibnamefont {Feng}}, \bibinfo
  {author} {\bibfnamefont {J.}~\bibnamefont {Chen}}, \bibinfo {author}
  {\bibfnamefont {Z.}~\bibnamefont {Wang}}, \bibinfo {author} {\bibfnamefont
  {H.}~\bibnamefont {Qiao}}, \bibinfo {author} {\bibfnamefont {W.}~\bibnamefont
  {Chen}}, \bibinfo {author} {\bibfnamefont {N.}~\bibnamefont {Zhao}}, \bibinfo
  {author} {\bibfnamefont {Z.}~\bibnamefont {Li}}, \bibinfo {author}
  {\bibfnamefont {J.}~\bibnamefont {Bu}}, \bibinfo {author} {\bibfnamefont
  {Z.}~\bibnamefont {Li}}, \bibinfo {author} {\bibfnamefont {Y.}~\bibnamefont
  {Liu}}, \bibinfo {author} {\bibfnamefont {Y.}~\bibnamefont {Zhao}},\ and\
  \bibinfo {author} {\bibfnamefont {D.}~\bibnamefont {Pei}},\ }in\ \href
  {https://doi.org/10.1145/3178876.3185996} {{\selectlanguage {en}\emph
  {\bibinfo {booktitle} {Proceedings of the 2018 {World} {Wide} {Web}
  {Conference} on {World} {Wide} {Web} - {WWW} '18}}}}\ (\bibinfo  {publisher}
  {ACM Press},\ \bibinfo {address} {Lyon, France},\ \bibinfo {year} {2018})\
  pp.\ \bibinfo {pages} {187--196}\BibitemShut {NoStop}%
\bibitem [{\citenamefont {Yao}\ \emph {et~al.}(2019)\citenamefont {Yao},
  \citenamefont {Liu}, \citenamefont {Zhang},\ and\ \citenamefont
  {Peng}}]{yao_unsupervised_2019}%
  \BibitemOpen
  \bibfield  {author} {\bibinfo {author} {\bibfnamefont {R.}~\bibnamefont
  {Yao}}, \bibinfo {author} {\bibfnamefont {C.}~\bibnamefont {Liu}}, \bibinfo
  {author} {\bibfnamefont {L.}~\bibnamefont {Zhang}},\ and\ \bibinfo {author}
  {\bibfnamefont {P.}~\bibnamefont {Peng}},\ }in\ \href
  {https://doi.org/10.1109/ICPHM.2019.8819434} {\emph {\bibinfo {booktitle}
  {2019 {IEEE} {International} {Conference} on {Prognostics} and {Health}
  {Management} ({ICPHM})}}}\ (\bibinfo  {publisher} {IEEE},\ \bibinfo {address}
  {San Francisco, CA, USA},\ \bibinfo {year} {2019})\ pp.\ \bibinfo {pages}
  {1--7}\BibitemShut {NoStop}%
\bibitem [{\citenamefont {Xu}\ \emph {et~al.}(2017)\citenamefont {Xu},
  \citenamefont {Sun}, \citenamefont {Deng},\ and\ \citenamefont
  {Tan}}]{xu_variational_2017}%
  \BibitemOpen
  \bibfield  {author} {\bibinfo {author} {\bibfnamefont {W.}~\bibnamefont
  {Xu}}, \bibinfo {author} {\bibfnamefont {H.}~\bibnamefont {Sun}}, \bibinfo
  {author} {\bibfnamefont {C.}~\bibnamefont {Deng}},\ and\ \bibinfo {author}
  {\bibfnamefont {Y.}~\bibnamefont {Tan}},\ }\bibfield  {journal} {\bibinfo
  {journal} {Proceedings of the AAAI Conference on Artificial Intelligence}\
  }\textbf {\bibinfo {volume} {31}},\ \href
  {https://doi.org/10.1609/aaai.v31i1.10966} {10.1609/aaai.v31i1.10966}
  (\bibinfo {year} {2017})\BibitemShut {NoStop}%
\bibitem [{\citenamefont {Mansour}\ \emph {et~al.}(2021)\citenamefont
  {Mansour}, \citenamefont {Escorcia-Gutierrez}, \citenamefont {Gamarra},
  \citenamefont {Gupta}, \citenamefont {Castillo},\ and\ \citenamefont
  {Kumar}}]{mansour_unsupervised_2021}%
  \BibitemOpen
  \bibfield  {author} {\bibinfo {author} {\bibfnamefont {R.~F.}\ \bibnamefont
  {Mansour}}, \bibinfo {author} {\bibfnamefont {J.}~\bibnamefont
  {Escorcia-Gutierrez}}, \bibinfo {author} {\bibfnamefont {M.}~\bibnamefont
  {Gamarra}}, \bibinfo {author} {\bibfnamefont {D.}~\bibnamefont {Gupta}},
  \bibinfo {author} {\bibfnamefont {O.}~\bibnamefont {Castillo}},\ and\
  \bibinfo {author} {\bibfnamefont {S.}~\bibnamefont {Kumar}},\ }\href
  {https://doi.org/10.1016/j.patrec.2021.08.018} {\bibfield  {journal}
  {\bibinfo  {journal} {Pattern Recognition Letters}\ }\textbf {\bibinfo
  {volume} {151}},\ \bibinfo {pages} {267} (\bibinfo {year}
  {2021})}\BibitemShut {NoStop}%
\bibitem [{\citenamefont {Wetzel}(2017)}]{wetzel_unsupervised_2017}%
  \BibitemOpen
  \bibfield  {author} {\bibinfo {author} {\bibfnamefont {S.~J.}\ \bibnamefont
  {Wetzel}},\ }\href {https://doi.org/10.1103/PhysRevE.96.022140} {\bibfield
  {journal} {\bibinfo  {journal} {Physical Review E}\ }\textbf {\bibinfo
  {volume} {96}},\ \bibinfo {pages} {022140} (\bibinfo {year}
  {2017})}\BibitemShut {NoStop}%
\bibitem [{\citenamefont {Khoshaman}\ \emph {et~al.}(2018)\citenamefont
  {Khoshaman}, \citenamefont {Vinci}, \citenamefont {Denis}, \citenamefont
  {Andriyash}, \citenamefont {Sadeghi},\ and\ \citenamefont
  {Amin}}]{khoshaman_quantum_2018}%
  \BibitemOpen
  \bibfield  {author} {\bibinfo {author} {\bibfnamefont {A.}~\bibnamefont
  {Khoshaman}}, \bibinfo {author} {\bibfnamefont {W.}~\bibnamefont {Vinci}},
  \bibinfo {author} {\bibfnamefont {B.}~\bibnamefont {Denis}}, \bibinfo
  {author} {\bibfnamefont {E.}~\bibnamefont {Andriyash}}, \bibinfo {author}
  {\bibfnamefont {H.}~\bibnamefont {Sadeghi}},\ and\ \bibinfo {author}
  {\bibfnamefont {M.~H.}\ \bibnamefont {Amin}},\ }\href
  {https://doi.org/10.1088/2058-9565/aada1f} {\bibfield  {journal} {\bibinfo
  {journal} {Quantum Science and Technology}\ }\textbf {\bibinfo {volume}
  {4}},\ \bibinfo {pages} {014001} (\bibinfo {year} {2018})}\BibitemShut
  {NoStop}%
\bibitem [{\citenamefont {Miles}\ \emph {et~al.}(2021)\citenamefont {Miles},
  \citenamefont {Carbone}, \citenamefont {Sturm}, \citenamefont {Lu},
  \citenamefont {Weichselbaum}, \citenamefont {Barros},\ and\ \citenamefont
  {Konik}}]{miles_machine_2021}%
  \BibitemOpen
  \bibfield  {author} {\bibinfo {author} {\bibfnamefont {C.}~\bibnamefont
  {Miles}}, \bibinfo {author} {\bibfnamefont {M.~R.}\ \bibnamefont {Carbone}},
  \bibinfo {author} {\bibfnamefont {E.~J.}\ \bibnamefont {Sturm}}, \bibinfo
  {author} {\bibfnamefont {D.}~\bibnamefont {Lu}}, \bibinfo {author}
  {\bibfnamefont {A.}~\bibnamefont {Weichselbaum}}, \bibinfo {author}
  {\bibfnamefont {K.}~\bibnamefont {Barros}},\ and\ \bibinfo {author}
  {\bibfnamefont {R.~M.}\ \bibnamefont {Konik}},\ }\href
  {https://doi.org/10.1103/PhysRevB.104.235111} {\bibfield  {journal} {\bibinfo
   {journal} {Physical Review B}\ }\textbf {\bibinfo {volume} {104}},\ \bibinfo
  {pages} {235111} (\bibinfo {year} {2021})}\BibitemShut {NoStop}%
\bibitem [{\citenamefont {Baul}\ \emph {et~al.}(2023)\citenamefont {Baul},
  \citenamefont {Walker}, \citenamefont {Moreno},\ and\ \citenamefont
  {Tam}}]{baul_application_2023}%
  \BibitemOpen
  \bibfield  {author} {\bibinfo {author} {\bibfnamefont {A.}~\bibnamefont
  {Baul}}, \bibinfo {author} {\bibfnamefont {N.}~\bibnamefont {Walker}},
  \bibinfo {author} {\bibfnamefont {J.}~\bibnamefont {Moreno}},\ and\ \bibinfo
  {author} {\bibfnamefont {K.-M.}\ \bibnamefont {Tam}},\ }\href
  {https://doi.org/10.1103/PhysRevE.107.045301} {\bibfield  {journal} {\bibinfo
   {journal} {Physical Review E}\ }\textbf {\bibinfo {volume} {107}},\ \bibinfo
  {pages} {045301} (\bibinfo {year} {2023})}\BibitemShut {NoStop}%
\bibitem [{\citenamefont {Goh}\ \emph {et~al.}(2022)\citenamefont {Goh},
  \citenamefont {Sheriffdeen}, \citenamefont {Wittmer},\ and\ \citenamefont
  {Bui-Thanh}}]{pmlr-v145-goh22a}%
  \BibitemOpen
  \bibfield  {author} {\bibinfo {author} {\bibfnamefont {H.}~\bibnamefont
  {Goh}}, \bibinfo {author} {\bibfnamefont {S.}~\bibnamefont {Sheriffdeen}},
  \bibinfo {author} {\bibfnamefont {J.}~\bibnamefont {Wittmer}},\ and\ \bibinfo
  {author} {\bibfnamefont {T.}~\bibnamefont {Bui-Thanh}},\ }in\ \href
  {https://proceedings.mlr.press/v145/goh22a.html} {\emph {\bibinfo {booktitle}
  {Proceedings of the 2nd mathematical and scientific machine learning
  conference}}},\ \bibinfo {series} {Proceedings of machine learning research},
  Vol.\ \bibinfo {volume} {145},\ \bibinfo {editor} {edited by\ \bibinfo
  {editor} {\bibfnamefont {J.}~\bibnamefont {Bruna}}, \bibinfo {editor}
  {\bibfnamefont {J.}~\bibnamefont {Hesthaven}},\ and\ \bibinfo {editor}
  {\bibfnamefont {L.}~\bibnamefont {Zdeborova}}}\ (\bibinfo  {publisher}
  {PMLR},\ \bibinfo {year} {2022})\ pp.\ \bibinfo {pages}
  {386--425}\BibitemShut {NoStop}%
\bibitem [{\citenamefont {He}\ \emph {et~al.}(2015)\citenamefont {He},
  \citenamefont {Zhang}, \citenamefont {Ren},\ and\ \citenamefont
  {Sun}}]{he_delving_2015}%
  \BibitemOpen
  \bibfield  {author} {\bibinfo {author} {\bibfnamefont {K.}~\bibnamefont
  {He}}, \bibinfo {author} {\bibfnamefont {X.}~\bibnamefont {Zhang}}, \bibinfo
  {author} {\bibfnamefont {S.}~\bibnamefont {Ren}},\ and\ \bibinfo {author}
  {\bibfnamefont {J.}~\bibnamefont {Sun}},\ }\href
  {http://arxiv.org/abs/1502.01852} {\bibinfo {title} {Delving {Deep} into
  {Rectifiers}: {Surpassing} {Human}-{Level} {Performance} on {ImageNet}
  {Classification}}} (\bibinfo {year} {2015}),\ \bibinfo {note}
  {arXiv:1502.01852 [cs]}\BibitemShut {NoStop}%
\bibitem [{\citenamefont {Kingma}\ and\ \citenamefont
  {Ba}(2017)}]{kingma_adam_2017}%
  \BibitemOpen
  \bibfield  {author} {\bibinfo {author} {\bibfnamefont {D.~P.}\ \bibnamefont
  {Kingma}}\ and\ \bibinfo {author} {\bibfnamefont {J.}~\bibnamefont {Ba}},\
  }\href {http://arxiv.org/abs/1412.6980} {\bibinfo {title} {Adam: {A} {Method}
  for {Stochastic} {Optimization}}} (\bibinfo {year} {2017}),\ \bibinfo {note}
  {arXiv:1412.6980 [cs]}\BibitemShut {NoStop}%
\bibitem [{\citenamefont {Reddi}\ \emph {et~al.}(2019)\citenamefont {Reddi},
  \citenamefont {Kale},\ and\ \citenamefont {Kumar}}]{reddi_convergence_2019}%
  \BibitemOpen
  \bibfield  {author} {\bibinfo {author} {\bibfnamefont {S.~J.}\ \bibnamefont
  {Reddi}}, \bibinfo {author} {\bibfnamefont {S.}~\bibnamefont {Kale}},\ and\
  \bibinfo {author} {\bibfnamefont {S.}~\bibnamefont {Kumar}},\ }\href
  {http://arxiv.org/abs/1904.09237} {\bibinfo {title} {On the {Convergence} of
  {Adam} and {Beyond}}} (\bibinfo {year} {2019}),\ \bibinfo {note}
  {arXiv:1904.09237 [cs, math, stat]}\BibitemShut {NoStop}%
\bibitem [{\citenamefont {Goodfellow}\ \emph {et~al.}(2016)\citenamefont
  {Goodfellow}, \citenamefont {Bengio},\ and\ \citenamefont
  {Courville}}]{goodfellow_deep_2016}%
  \BibitemOpen
  \bibfield  {author} {\bibinfo {author} {\bibfnamefont {I.}~\bibnamefont
  {Goodfellow}}, \bibinfo {author} {\bibfnamefont {Y.}~\bibnamefont {Bengio}},\
  and\ \bibinfo {author} {\bibfnamefont {A.}~\bibnamefont {Courville}},\
  }\href@noop {} {\emph {\bibinfo {title} {Deep learning}}},\ Adaptive
  computation and machine learning\ (\bibinfo  {publisher} {The MIT Press},\
  \bibinfo {address} {Cambridge, Massachusetts},\ \bibinfo {year}
  {2016})\BibitemShut {NoStop}%
\bibitem [{\citenamefont {Boehnke}\ \emph {et~al.}(2011)\citenamefont
  {Boehnke}, \citenamefont {Hafermann}, \citenamefont {Ferrero}, \citenamefont
  {Lechermann},\ and\ \citenamefont {Parcollet}}]{boehnke_orthogonal_2011}%
  \BibitemOpen
  \bibfield  {author} {\bibinfo {author} {\bibfnamefont {L.}~\bibnamefont
  {Boehnke}}, \bibinfo {author} {\bibfnamefont {H.}~\bibnamefont {Hafermann}},
  \bibinfo {author} {\bibfnamefont {M.}~\bibnamefont {Ferrero}}, \bibinfo
  {author} {\bibfnamefont {F.}~\bibnamefont {Lechermann}},\ and\ \bibinfo
  {author} {\bibfnamefont {O.}~\bibnamefont {Parcollet}},\ }\href
  {https://doi.org/10.1103/PhysRevB.84.075145} {\bibfield  {journal} {\bibinfo
  {journal} {Physical Review B}\ }\textbf {\bibinfo {volume} {84}},\ \bibinfo
  {pages} {075145} (\bibinfo {year} {2011})}\BibitemShut {NoStop}%
\bibitem [{\citenamefont {White}(1991)}]{white_spectral_1991}%
  \BibitemOpen
  \bibfield  {author} {\bibinfo {author} {\bibfnamefont {S.~R.}\ \bibnamefont
  {White}},\ }\href {https://doi.org/10.1103/PhysRevB.44.4670} {\bibfield
  {journal} {\bibinfo  {journal} {Physical Review B}\ }\textbf {\bibinfo
  {volume} {44}},\ \bibinfo {pages} {4670} (\bibinfo {year}
  {1991})}\BibitemShut {NoStop}%
\bibitem [{\citenamefont {Claesen}\ and\ \citenamefont
  {De~Moor}(2015)}]{claesen_hyperparameter_2015}%
  \BibitemOpen
  \bibfield  {author} {\bibinfo {author} {\bibfnamefont {M.}~\bibnamefont
  {Claesen}}\ and\ \bibinfo {author} {\bibfnamefont {B.}~\bibnamefont
  {De~Moor}},\ }\href {https://arxiv.org/abs/1502.02127} {\bibinfo {title}
  {Hyperparameter {Search} in {Machine} {Learning}}} (\bibinfo {year} {2015}),\
  \bibinfo {note} {arXiv:1502.02127 [cs.LG]}\BibitemShut {NoStop}%
\bibitem [{\citenamefont {Ganaie}\ \emph {et~al.}(2022)\citenamefont {Ganaie},
  \citenamefont {Hu}, \citenamefont {Malik}, \citenamefont {Tanveer},\ and\
  \citenamefont {Suganthan}}]{ganaie_ensemble_2022}%
  \BibitemOpen
  \bibfield  {author} {\bibinfo {author} {\bibfnamefont {M.}~\bibnamefont
  {Ganaie}}, \bibinfo {author} {\bibfnamefont {M.}~\bibnamefont {Hu}}, \bibinfo
  {author} {\bibfnamefont {A.}~\bibnamefont {Malik}}, \bibinfo {author}
  {\bibfnamefont {M.}~\bibnamefont {Tanveer}},\ and\ \bibinfo {author}
  {\bibfnamefont {P.}~\bibnamefont {Suganthan}},\ }\href
  {https://doi.org/10.1016/j.engappai.2022.105151} {\bibfield  {journal}
  {\bibinfo  {journal} {Engineering Applications of Artificial Intelligence}\
  }\textbf {\bibinfo {volume} {115}},\ \bibinfo {pages} {105151} (\bibinfo
  {year} {2022})}\BibitemShut {NoStop}%
\bibitem [{\citenamefont {Jarrell}\ and\ \citenamefont
  {Gubernatis}(1996)}]{jarrell_bayesian_1996}%
  \BibitemOpen
  \bibfield  {author} {\bibinfo {author} {\bibfnamefont {M.}~\bibnamefont
  {Jarrell}}\ and\ \bibinfo {author} {\bibfnamefont {J.}~\bibnamefont
  {Gubernatis}},\ }\href {https://doi.org/10.1016/0370-1573(95)00074-7}
  {\bibfield  {journal} {\bibinfo  {journal} {Physics Reports}\ }\textbf
  {\bibinfo {volume} {269}},\ \bibinfo {pages} {133} (\bibinfo {year}
  {1996})}\BibitemShut {NoStop}%
\bibitem [{\citenamefont {Levy}\ \emph {et~al.}(2017)\citenamefont {Levy},
  \citenamefont {LeBlanc},\ and\ \citenamefont
  {Gull}}]{levy_implementation_2017}%
  \BibitemOpen
  \bibfield  {author} {\bibinfo {author} {\bibfnamefont {R.}~\bibnamefont
  {Levy}}, \bibinfo {author} {\bibfnamefont {J.~P.~F.}\ \bibnamefont
  {LeBlanc}},\ and\ \bibinfo {author} {\bibfnamefont {E.}~\bibnamefont
  {Gull}},\ }\href {https://doi.org/10.17632/RF3P4PSDHS.1} {\bibinfo {title}
  {Implementation of the maximum entropy method for analytic continuation}}
  (\bibinfo {year} {2017}),\ \bibinfo {note}
  {10.17632/RF3P4PSDHS.1}\BibitemShut {NoStop}%
\bibitem [{\citenamefont {Randeria}\ \emph {et~al.}(2005)\citenamefont
  {Randeria}, \citenamefont {Sensarma}, \citenamefont {Trivedi},\ and\
  \citenamefont {Zhang}}]{randeria_particle-hole_2005}%
  \BibitemOpen
  \bibfield  {author} {\bibinfo {author} {\bibfnamefont {M.}~\bibnamefont
  {Randeria}}, \bibinfo {author} {\bibfnamefont {R.}~\bibnamefont {Sensarma}},
  \bibinfo {author} {\bibfnamefont {N.}~\bibnamefont {Trivedi}},\ and\ \bibinfo
  {author} {\bibfnamefont {F.-C.}\ \bibnamefont {Zhang}},\ }\href
  {https://doi.org/10.1103/PhysRevLett.95.137001} {\bibfield  {journal}
  {\bibinfo  {journal} {Physical Review Letters}\ }\textbf {\bibinfo {volume}
  {95}},\ \bibinfo {pages} {137001} (\bibinfo {year} {2005})}\BibitemShut
  {NoStop}%
\bibitem [{\citenamefont {Bai}\ \emph {et~al.}(2007)\citenamefont {Bai},
  \citenamefont {Chen}, \citenamefont {Scalettar},\ and\ \citenamefont
  {Yamazaki}}]{bai_robust_2007}%
  \BibitemOpen
  \bibfield  {author} {\bibinfo {author} {\bibfnamefont {Z.}~\bibnamefont
  {Bai}}, \bibinfo {author} {\bibfnamefont {W.}~\bibnamefont {Chen}}, \bibinfo
  {author} {\bibfnamefont {R.}~\bibnamefont {Scalettar}},\ and\ \bibinfo
  {author} {\bibfnamefont {I.}~\bibnamefont {Yamazaki}},\ }\href@noop {}
  {\bibfield  {journal} {\bibinfo  {journal} {Proceedings of the 4th
  International Congress of Chinese Mathematician (ICCM), Edited by L. Ji, K.
  Liu, L. Yang, S.-T. Yau}\ }\textbf {\bibinfo {volume} {III}},\ \bibinfo
  {pages} {253} (\bibinfo {year} {2007})}\BibitemShut {NoStop}%
\end{thebibliography}%


\begin{thebibliography}{0}%
\makeatletter
\providecommand \@ifxundefined [1]{%
 \@ifx{#1\undefined}
}%
\providecommand \@ifnum [1]{%
 \ifnum #1\expandafter \@firstoftwo
 \else \expandafter \@secondoftwo
 \fi
}%
\providecommand \@ifx [1]{%
 \ifx #1\expandafter \@firstoftwo
 \else \expandafter \@secondoftwo
 \fi
}%
\providecommand \natexlab [1]{#1}%
\providecommand \enquote  [1]{``#1''}%
\providecommand \bibnamefont  [1]{#1}%
\providecommand \bibfnamefont [1]{#1}%
\providecommand \citenamefont [1]{#1}%
\providecommand \href@noop [0]{\@secondoftwo}%
\providecommand \href [0]{\begingroup \@sanitize@url \@href}%
\providecommand \@href[1]{\@@startlink{#1}\@@href}%
\providecommand \@@href[1]{\endgroup#1\@@endlink}%
\providecommand \@sanitize@url [0]{\catcode `\\12\catcode `\$12\catcode
  `\&12\catcode `\#12\catcode `\^12\catcode `\_12\catcode `\%12\relax}%
\providecommand \@@startlink[1]{}%
\providecommand \@@endlink[0]{}%
\providecommand \url  [0]{\begingroup\@sanitize@url \@url }%
\providecommand \@url [1]{\endgroup\@href {#1}{\urlprefix }}%
\providecommand \urlprefix  [0]{URL }%
\providecommand \Eprint [0]{\href }%
\providecommand \doibase [0]{https://doi.org/}%
\providecommand \selectlanguage [0]{\@gobble}%
\providecommand \bibinfo  [0]{\@secondoftwo}%
\providecommand \bibfield  [0]{\@secondoftwo}%
\providecommand \translation [1]{[#1]}%
\providecommand \BibitemOpen [0]{}%
\providecommand \bibitemStop [0]{}%
\providecommand \bibitemNoStop [0]{.\EOS\space}%
\providecommand \EOS [0]{\spacefactor3000\relax}%
\providecommand \BibitemShut  [1]{\csname bibitem#1\endcsname}%
\let\auto@bib@innerbib\@empty
\end{thebibliography}%
\putbib
\end{bibunit}

\clearpage
\setcounter{equation}{0}
\setcounter{figure}{0}
\setcounter{table}{0}

\makeatletter
\renewcommand{\theequation}{S\arabic{equation}}
\renewcommand{\thefigure}{S\arabic{figure}}
\clearpage
\onecolumngrid
\begin{center}
{\large Supplemental Material for\\[2mm] \bf Autoencoder-based analytic continuation method\\[1mm] for strongly correlated quantum
systems}\\[3mm]
Maksymilian Kliczkowski,$^1$ Lauren Keyes,$^2$ Sayantan Roy,$^2$ Thereza~Paiva,$^3$\\ Mohit Randeria,$^2$ Nandini Trivedi$^2$, and Maciej~M.~Ma\'ska$^1$\\[2mm]
{\small $^1${\it Institute of Theoretical Physics, Wroc{\l}aw University of Science and Technology, 50-370 Wroc{\l}aw, Poland}\\[0mm]
$^2${\it Department of Physics, The Ohio State University, Columbus, Ohio 43210, USA}}\\
$^3${\it Instituto de F\'isica, Universidade Federal do Rio de Janeiro Cx.P. 68.528, 21941-972 Rio de Janeiro RJ, Brazil}\\
\end{center}
\setcounter{page}{1}
\vspace*{5mm}

\begin{bibunit}
\twocolumngrid

\subsection{Model Hamiltonian\label{sec:Hamiltonian}}
We demonstrate the proposed AE approach in application to QMC data generated for the Fermi-Hubbard model. In particle-hole symmetric form it is described by the following Hamiltonian,
\begin{align}\label{eq:hubbard}
\hat{H} =& -t \sum _{\langle i,j\rangle\sigma}
 \left(\hat{c}^\dag_{i\sigma}\hat{c}_{j\sigma} + \mathrm{H.c.}\right)  \nonumber \\
& + U \sum_i \left(\hat{n}_{i\uparrow}-\frac{1}{2}\right)\left(\hat{n}_{i\downarrow}-\frac{1}{2}\right) - \mu \hat{N},
\end{align}
where $\hat{c}_{i\sigma}^\dag$ ($\hat{c}_{i\sigma}$) denotes fermionic creation (annihilation) operators, $t$ is the hopping integral, and $U$ denotes the on-site Coulomb repulsion. Fermionic number operators are defined as $\hat{n}_{i\sigma}=\hat{c}^\dag_{i\sigma}\hat{c}_{i\sigma}, \hat{n}_i = \hat{n}_{i\uparrow}+\hat{n}_{i\downarrow}, \hat{N} = \sum _i \hat{n}_i$. 

\subsection{Details of the Autoencoder Neural Network and its training and testing\label{sec:NNdetails}}
In this section, we describe the methods we use to perform the \textit{analytic continuation} for both artificially generated Green's functions and those obtained from the QMC simulation. We briefly describe the NN architecture in terms of technical details of the trainable part. We begin the discussion by providing the form of the input data. The SFs presented in the scope of this work have two separate origins. We introduce both of them in the following subsections. Namely, we establish the procedure of generating the \textit{pretraining} SFs dataset in order to validly perform \textit{training} on the Green's functions obtained from the QMC.
\subsubsection{Pretraining procedure}
\begin{figure}[!ht]
\centerline{\includegraphics[width=0.390\textwidth]{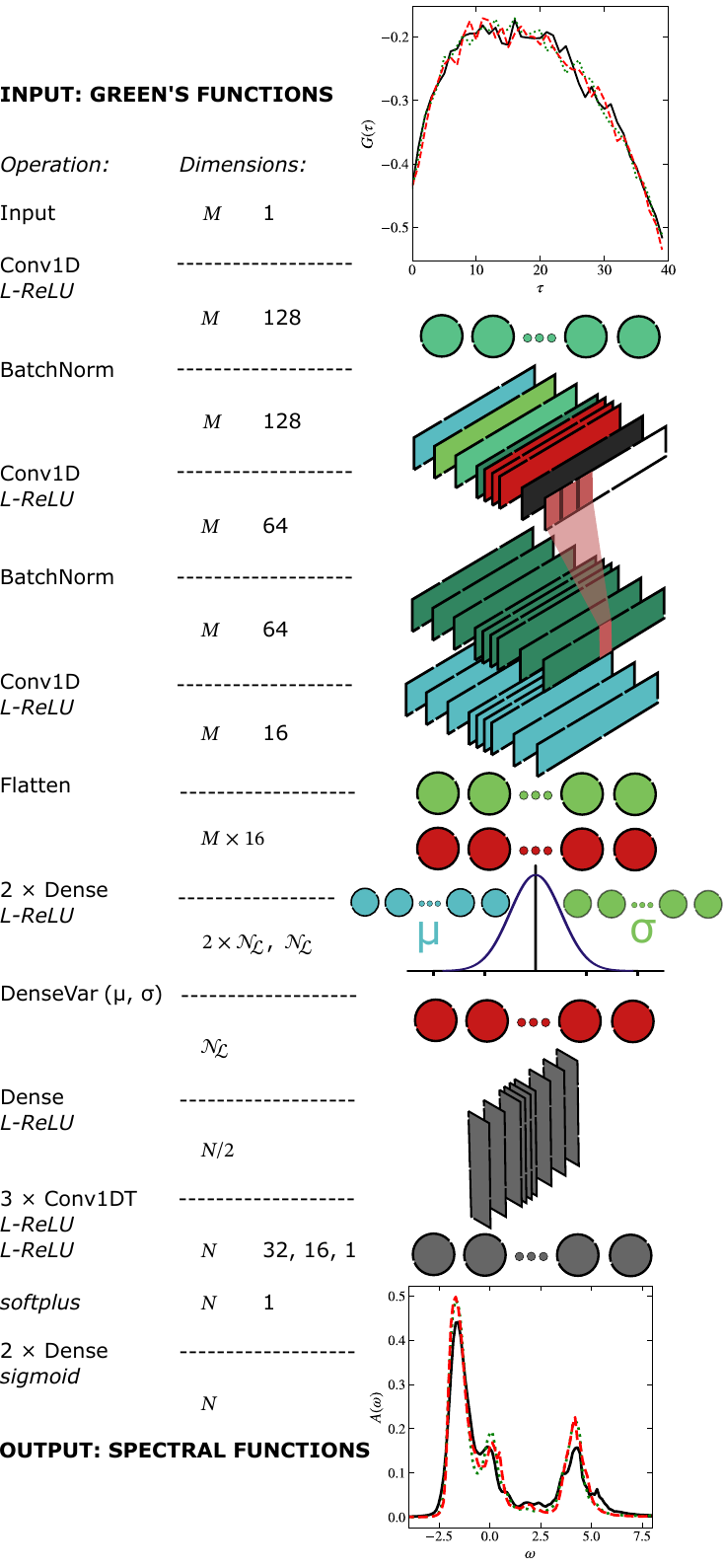}}
\caption{An example architecture of the \raisebox{-0.7mm}{\includegraphics*[width=5mm]{FinalFig/EN.pdf}} part of the NN with $N$ frequency units and $M$ imaginary Matsubara times. Layers are shown graphically with their corresponding dimensions. Reshape layers have been omitted for simplicity. The layer names are according to Keras API \cite{chollet_keras_2015}. The network was implemented in the TensorFlow framework \cite{tensorflow_developers_tensorflow_2023}. \label{fig:architecture}}
\end{figure}
During the \textit{pretraining}, defined as a stage of training the initial weights of the \raisebox{-0.7mm}{\includegraphics*[width=5mm]{FinalFig/EN.pdf}}, we introduce a substantial number of ``artificial'' SFs $\{\vec{\mathcal{A}}^n\}$. These are modeled to feature multiple Gaussian peaks randomly located at various frequencies, each with random weights. Subsequently, we solve the forward problem, as outlined in Eq.~(\ref{eq:direct}), to compute corresponding Green's functions $\{\vec{\mathcal{G}}^n\}$ at a specified inverse temperature $\beta$. To ensure that these SFs are physically meaningful, we adhere to the recently proposed generation procedure by Zhang et al. \cite{zhang_training_2022}, a method that has been shown to effectively optimize the weights of the NN. 
In order to perform the \textit{pretraining} process, we employ pairs of $(\vec{\mathcal{G}}^n, \vec{\mathcal{A}}_0^n)$, as elaborated in the main text. We introduce statistical noise as random normal variables $\mathcal{N}(0, \sigma^2)$ for each imaginary time point $\tau$ within $\{\vec{\mathcal{G}}^n\}$'s. The variance, denoted as $\sigma$, characterizing the Green's function, which is typically obtained by statistical means from the QMC, serves as a metric in Fig.~\ref{fig:errors} to quantify the resilience of the AE and MaxEnt approaches in the presence of deviations resulting from finite numerical simulations. 


\subsubsection{Network architecture}
In this Letter, we propose the \textit{autoencoder} approach to \textit{analytic continuation}, as described in Eq.~(\ref{eq:autoencoder}). While the number of imaginary times, denoted as $M$ ($M=40$ in most cases), in Green's functions, and $N=250, 1000$ frequencies in the SFs, impose constraints on the outer layers of the \raisebox{-0.7mm}{\includegraphics*[width=5mm]{FinalFig/EN.pdf}} part of the AE (specifically, the \textit{input} and the \textit{output}), the internal structure of the network architecture is by no means restricted. Moreover, the \raisebox{-0.7mm}{\includegraphics*[width=5mm]{FinalFig/DE.pdf}} part is strictly determined by the Fredholm integral [Eq.~(\ref{eq:direct})] and can be implemented as a fully connected, non-trainable dense layer, featuring weights defined by Eq.~(\ref{eq:kernel}) and zero bias. The presence of statistical noise originating from QMC simulations suggests that it is reasonable to utilize network architectures that are capable of discriminating between noise and relevant data. In this context, we choose to implement a variational autoencoder (VAE) type network. 
VAEs represent a class of NNs that are characterized by their generative capabilities \cite{kingma_introduction_2019, cinelli_variational_2021}. That is, they model the distribution of the true data generation by introducing a latent dimension. In this context, there resides a single VAE within \raisebox{-0.7mm}{\includegraphics*[width=5mm]{FinalFig/EN.pdf}}, since the true \raisebox{-0.7mm}{\includegraphics*[width=5mm]{FinalFig/DE.pdf}} is determined a priori. VAEs have been shown to be effective across a spectrum of optimization problems, including anomaly detection \cite{xu_unsupervised_2018, yao_unsupervised_2019}, text classification \cite{xu_variational_2017}, illness diagnosis and classification \cite{mansour_unsupervised_2021}, and physical problems \cite{wetzel_unsupervised_2017, khoshaman_quantum_2018,  miles_machine_2021, baul_application_2023}. The ability of VAEs to faithfully characterize each element of the input set through the latent probability distribution makes them a potentially powerful tool to address inverse problems \cite{pmlr-v145-goh22a}. 

In the initial part of \raisebox{-0.7mm}{\includegraphics*[width=5mm]{FinalFig/EN.pdf}}, we opt for the utilization of 1-dimensional convolutional (Conv1D) layers featuring a kernel of size 3 and a stride of size 1. The application of one-dimensional kernels allows the network to grasp the monotonic patterns within the input data. This further contributes to diminishing the impact of noise on the Green's functions. We adopt the initialization of the weights for the \textit{pretraining} procedure akin to He et al. \cite{he_delving_2015}. Batch normalization is applied between Conv1D layers with activation as a \textit{leaky} implementation of rectifier linear units (L-ReLU) in order to avoid overfitting to a single noise realization. The latent space representation comprises $\mathcal{N_L} = 120$ fully-connected neurons. To transform the encoded values within the latent dimension, we employ multiple transposed Conv1D (Conv1DT) layers. The final output of \raisebox{-0.7mm}{\includegraphics*[width=5mm]{FinalFig/EN.pdf}} consists of two fully connected feed-forward layers, each comprising $N$ units, with the sigmoid activation function facilitating the normalization condition. 
Both \textit{pretraining} and \textit{training} procedures are executed using the Adam optimizer \cite{kingma_adam_2017} with \textit{ams-grad} enhancement \cite{reddi_convergence_2019}. For efficient learning, we use the early stopping condition on the validation dataset \cite{goodfellow_deep_2016}. An example of the architecture used for the generation of QMC SFs for the Hubbard model from Eq.~(\ref{eq:hubbard}) is visualized in Fig.~\ref{fig:architecture}. In principle, the architecture allows for performing the analytic continuation for various momenta $\vec{k}$ simultaneously. As we wanted to make the presentation more accessible, we only experimented with this idea, but plan to return to it for future development. Additionally, to further control over the noise level and input dimensionality, one can approximate the input Green's functions using Legendre polynomials, with the flexibility to make adjustments by manipulating their order~\cite{boehnke_orthogonal_2011}. Although we do not currently employ this technique, it can be easily applied within our framework. 

\subsubsection{Testing procedure}

We illustrate the procedure that we used to compare the effectiveness of the MaxEnt and AE approaches. Panel (a) in Fig.~\ref{fig:greensandspectrals1} in the main text shows an example of a ``artificial'' SF. Then, Eq.~\eqref{eq:direct} is used to generate the corresponding Green's function. In the next step, a series of noisy Green's functions ${\cal G}_\sigma$ is generated by introducing random disorder of various magnitudes $\sigma$. Examples are shown in panel (c). An increase in the noise makes the inverse problem harder to optimize. This can be seen in panels (b) and (d), which show the results of the analytic continuation performed with the help of the MaxEnt and AE methods, respectively.  As expected, reducing the noise level improves the ability of both methods to find the corresponding spectral function more accurately from the noisy set $\{\vec{\mathcal{G}}_\sigma^n\}$. However, one can notice there that the AE method gives more accurate results for large and intermediate noise levels. This confirms that shorter QMC runs are needed for AE to obtain the same accuracy as for the MaxEnt method.
\subsection{Custom regularization}
In this subsection we elaborate on defining the regularization for the NN. The \raisebox{-0.7mm}{\includegraphics*[width=5mm]{FinalFig/EN.pdf}} part of the AE has to solve the problem of transforming the input function into a function which, when integrated, gives the same input function. The transformation is determined by weights and biases, which are adjusted during the training stage. Since this is an ill-posed problem, it is reasonable to expect that the final values of these weights and biases will not be optimal, resulting in divergence in the output of AE from the input. 

Training is an iterative process that starts with random (or pretrained) initial values of the weights and biases, and therefore the final parameters of the neural network can also depend on initialization. To minimize this dependence, we employ a technique to improve the results of a {\it single initialization}. Since integration ``smooths out'' irregularities in the integrand function, the output of the \raisebox{-0.7mm}{\includegraphics*[width=5mm]{FinalFig/DE.pdf}} part is highly insensitive to high-frequency oscillations or noise in the output of the encoder part. Consequently, to extract the essential features of SFs, we incorporate regularization methods commonly used in inverse problems. These regularization techniques help stabilize and improve the reconstruction process. To this end, the total cost function is defined as
\begin{equation}
    \label{eq:cost}
    \mathcal{L} = \chi^2(\vec{G},\vec{G}') + \sum_m \alpha _m \eta ^m (\vec{G},\vec{G}';\vec{A}),
\end{equation}
where the regularization strength $\alpha _m$ describes the significance of the specific regularization function $\eta ^m (\vec{G}, \vec{G}';\vec{A})$. The interplay between different penalty terms enables a more meaningful solution, especially in situations where the transformation is inherently ambiguous. In the MaxEnt procedure, as described in the preceding section, the loss $\mathcal{L}$ is accompanied by an \textit{entropy} penalty, which is used to minimize spurious correlations between the data by introducing a default reference model. Typically, a stronger regularization leads to a smoother output during the procedure. A notable advantage of this approach lies in its flexibility to incorporate arbitrary regularization terms. In particular, we consider physically relevant terms that penalize deviations from the \textit{sum rules}, which can be analytically calculated for the Hubbard model up to the second moment \cite{white_spectral_1991}. They are defined as
\begin{subequations}\label{eq:sum_rules}
    \begin{align}
        m_0 & = \int_{-\infty}^{\infty}d\omega A(k,\omega) = 1, \\
        m_1 & = \int_{-\infty}^{\infty}d\omega\: \omega A(k,\omega) = \varepsilon_k-\mu+\frac{U}{2}( n  -1), \\
        m_2 & = \int_{-\infty}^{\infty}d\omega\: \omega ^2 A(k,\omega) = \left(\varepsilon_k-\mu-\frac{U}{2}\right)^2 \nonumber \\
        &       + U\left(\varepsilon_k-\mu-\frac{U}{2}\right) n +\frac{1}{2}U^2 n , 
    \end{align}
\end{subequations}
where $\varepsilon_k = -2t(\cos{k_x}+\cos{k_y})$ is the tight binding dispersion on the square lattice with only nearest neighbor hopping, $t$. For some of our results, we have employed hyper-parameter optimization techniques such as grid search or Bayesian search \cite{claesen_hyperparameter_2015}. It should be noted that there may be potential improvements in the quality of the results by additionally applying \textit{ensemble learning} techniques \cite{ganaie_ensemble_2022}.
\subsection{MaxEnt procedure} \label{maxent procedure}

The maximum entropy method for analytic continuation uses Bayesian principles to identify the spectral function which, given some Green's function data, minimizes the functional
\begin{equation}
    Q = \frac12 \chi^2 -\alpha S[A(\omega)].
    \label{Q eqn}
\end{equation}
\noindent
$Q$ is the canonical symbol used for this functional, and its role is analogous to the loss $\mathcal{L}$ in machine learning techniques. Minimizing $Q$ is equivalent to performing a chi-squared fitting, regularized by the Shannon entropy term $S[A(\omega)]=-\int d\omega\ A(\omega)\ln[A(\omega)/d(\omega)]$. The function $d(\omega)$ is known as the default model, which must be chosen using prior knowledge about the physical nature of the SF. The parameter $\alpha$ controls the relative strength between the $\chi ^2$ and entropy terms in $Q$. If $\alpha >> 1$, then the MaxEnt method would give $d(\omega)$ as the solution for the SF \cite{jarrell_bayesian_1996}.

The primary hurdles in using MaxEnt are the identification of the appropriate value of $\alpha$ and the appropriate choice of $d(\omega)$. There have been various approaches to choosing $\alpha$, including Bryan's method, which finds the SF which minimizes Eq. \eqref{Q eqn} for a range of $\alpha$ values, then gives an average SF, weighted by the probability of each $\alpha$ \cite{jarrell_bayesian_1996}. It is difficult to choose a default model, since the structure of the spectral function is {\it a priori} unknown. One can use general properties, such as whether the system is in a conducting or insulating state, but even these basic properties may be unknown when studying a model with an incomplete phase diagram.

In this paper, we use the MaxEnt code implemented by Levy \cite{levy_implementation_2017}. We treat $\alpha$ with Bryan's method, using 60 equally spaced $\alpha$ values in the range $0.1\leq\alpha\leq 20$. The SF is calculated with 250 equally spaced frequencies, ranging from $-15<\omega t < 15$, where $t$ sets the scale of the hopping energy. The choice of default model is described in the sections below.

\subsubsection{MaxEnt for artificial spectral functions}
We perform the MaxEnt procedure for the Green's functions (Fig. \ref{fig:errors}), obtained from artificially generated spectral functions using a uniform default model. This is due to the fact that there is no physical intuition allowing us to choose a different model.
\subsubsection{MaxEnt at half-filling}
For the Green's functions produced by DQMC at $U/t=8$, $\mu = 0$ and $\beta=0.5$ (Fig. \ref{fig:conver}), the default model is chosen to consist of two Gaussian peaks, centered at $\omega t = \pm 5$ and both with a standard deviation of $1$. This is due to the repulsive Hubbard model likely being in a gapped state for these parameters. 
\subsubsection{MaxEnt away from half-filling}

A consequence of doping the repulsive Hubbard model away from the Mott insulating limits is an asymmetry in the local density of states (LDOS) ~\cite{randeria_particle-hole_2005}. With increasing particle doping, the LDOS spectrum shifts to the left, to satisfy the sum rules.  Since the choice of default model is ambiguous, we perform the analytic continuation with uniform, Lorentzian and Gaussian default models, while varying the width of the last two models. Choice of the optimum spectral functions is based on which default model reproduces the first three moments of the spectral function the closest, as defined in Eq.~(\ref{eq:sum_rules}).
\subsection{DQMC}
The Green's functions shown in this Letter are produced with the ``bandmott'' version of the determinant quantum Monte Carlo code~\cite{bai_robust_2007}. The simulation is done on a square lattice of side length $L_x=L_y=16$. Because of symmetries of the lattice, only the lower trianglular half of the first quadrant is computed. The inverse temperature is kept at $\beta = 2t$ and the Hubbard interaction at $U/t=8$. The Trotter error is fixed at $\Delta\tau = 0.05$.

\subsubsection{DQMC at half-filling}
All of the the Green's functions shown in the main text were calculated at half-filling. The results were obtained using 1000 warmup sweeps, performed before the measurement sweeps. Every next measurement is taken after 10 sweeps through the auxiliary fields.
\begin{figure}[!ht]
\centerline{\includegraphics[width=0.48\textwidth]{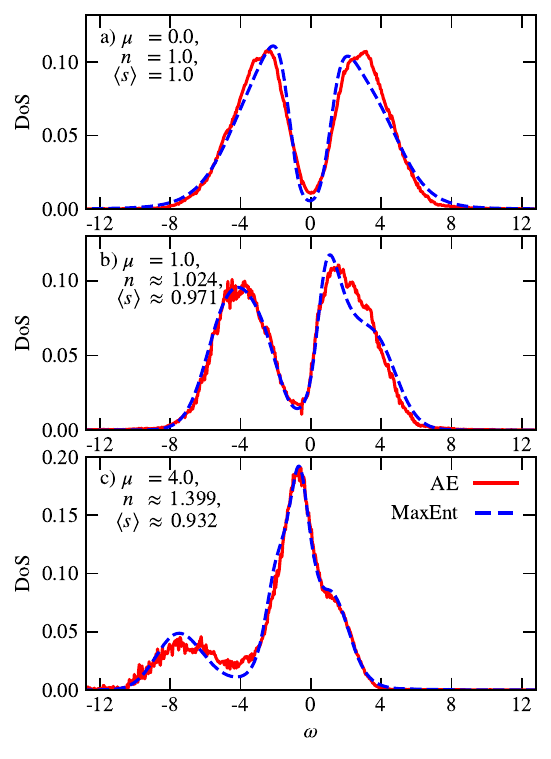}}
\caption{Density of states calculated with the help of the AE (red lines) and MaxEnt (blue lines) methods for the Hubbard model from Eq.~(\ref{eq:hubbard}) at $U=8$ on a $16 \times 16$ square lattice. The procedures were carried out in three different particle fillings (a) $n=1.0$, (b) $n\approx1.024$ and (c) $n\approx 1.399$.   \label{fig:DOS}}
\end{figure}
With the exception of Fig.~\ref{fig:conver}, all data is produced from $40$ measurements and 100 different trials ($m_s=4\times 10^3$ measurements in total). A ``trial'' refers to one instance of the DQMC system produced by a random initial seed. 

In Fig.~\ref{fig:conver}, the number of measurements is varied, but all other parameters are kept the same. Separate trials are produced by different random seeds. The largest error bars, labeled ``$\sigma=2.5$E$-2$'' in the figure, result from $m_s = 15$ measurements on one trial, ``$\sigma=2.6$E$-3$'' is the result of $m_s = 40$ measurements on one trial, ``$\sigma=2.8$E$-4$'' is produced by performing 40 measurements on 100 different trials ($m_s=4\times 10^3$ measurements in total), and ``$\sigma = 8.8$E$-5$'' is produced by performing 40 measurements on 1002 different trials ($m_s \approx 4 \times 10^4$ measurements in total). The $\vec{A}^\infty _{M, A}$ is obtained from the Green's functions originating from 40 measurements on 3000 different random seeds so ($m_s = 1.2 \times 10^5$ measurements in total). The error bar on each $G_{\vec{k}}(\tau)$ is given by the standard error over the measurements.
\subsubsection{DQMC away from half-filling}

The QMC simulations away from half-filling were performed with 2000 warmup sweeps and $m_s=5000$ measurement sweeps. Measurement of the Green's functions was done after every 10 measurement sweeps. The Green's functions were averaged over runs from 20 independent initial configurations of the Hubbard-Stratonovich fields. The MaxEnt was performed at dopings outlined in Tab.~\ref{tab:app:doping}. 
\begin{table}[!ht]
\centering
\begin{tabular}{|c|c|c|}
\hline
     $\mu$ & $\langle s \rangle$ & $n$ \\
     \hline
     1.00 &0.9714 & 1.0237   \\
     \hline
     4.00 &0.9324 & 1.3988 \\
     \hline   
\end{tabular}
\caption{Average particle dopings $n$ and average sign $\langle s \rangle$ resulting from the DQMC of the Hubbard model from Eq.~\eqref{eq:hubbard} away from half filling ($\mu \neq 0$).\label{tab:app:doping}}
\end{table}
For the Lorentzian default model, widths of $\Gamma \in \{0.5,1.0,2.0\}$ were used, whereas for Gaussian default model, widths of $\sigma \in \{1.0,2.0\}$ were used. Out of these, the Gaussian default model with $\sigma = 2.0$ gave the closest results to the sum rules [Eq.~(\ref{eq:sum_rules})].

We compare both AE and MaxEnt methods for different dopings in Fig.~\ref{fig:DOS} by showing the Density of States (DoS), i.e., the SFs summed over all momenta, for different electron concentrations. The validity of the analytically continued spectral functions used in Fig.~\ref{fig:DOS} was checked by calculating the imaginary time Green's function by the forward integration [Eq.~(\ref{eq:direct})], which matched the Green's functions obtained directly from QMC simulations within at most 1\% error. 

\putbib
\end{bibunit}
\end{document}